\documentclass[aps, twocolumn, superscript address, nobalancelastpage]{revtex4-1}
\usepackage[colorlinks=true,citecolor=blue,linkcolor=blue,urlcolor=red]{hyperref}
\usepackage{amsmath}
\usepackage[sort&compress]{natbib}
\usepackage{graphicx,times}
\usepackage{color}
\usepackage{sansmath}
\usepackage{amssymb}
\usepackage{subeqnarray}
\usepackage{diagbox}
\usepackage{makecell}
\usepackage{tabularx}

\newcommand{\ket}[1]{\ensuremath{\vert{#1}\rangle}}
\newcommand{\bra}[1]{\ensuremath{\langle{#1}\vert}}
\newcommand{\abs}[1]{\ensuremath{\left|{#1} \right|}}
\newcommand{\msf}[1]{\mathsf{#1}}
\newcommand{\mcl}[1]{\mathcal{#1}}

\newcommand{\parent}[1]{\left( #1 \right)}
\newcommand{\mathbbm}[1]{\text{\usefont{U}{bbm}{m}{n}#1}}

\newcommand{\cor}[1]{\left[ #1 \right]}
\newcommand{\bsb}[1]{\boldsymbol{#1}}

\begin{document}
\title{Chiral states and nonreciprocal phases in a Josephson junction ring}
\author{R. Asensio-Perea}
\affiliation{Department of Physical Chemistry, University of the Basque Country UPV/EHU, Apartado 644, E-48080 Bilbao, Spain}
\author{A. Parra-Rodriguez}
\affiliation{Department of Physics, University of the Basque Country UPV/EHU, Apartado 644, 48080 Bilbao, Spain}
\author{G. Kirchmair}
\affiliation{Institute for Quantum Optics and Quantum Information of the Austrian Academy of Sciences, A-6020 Innsbruck, Austria}
\affiliation{Institute for Experimental Physics, University of Innsbruck, A-6020 Innsbruck, Austria}
\author{E. Solano}
\affiliation{Department of Physical Chemistry, University of the Basque Country UPV/EHU, Apartado 644, E-48080 Bilbao, Spain}
\affiliation{IQM, Nymphenburgerstr. 86, 80636  Munich, Germany}
\affiliation{International Center of Quantum Artificial Intelligence for Science and Technology (QuArtist) \\ and Department of Physics, Shanghai University, 200444 Shanghai, China}
\affiliation{IKERBASQUE, Basque Foundation for Science, Plaza Euskadi 5, 48009 Bilbao, Spain}
\author{E. Rico}
\affiliation{Department of Physical Chemistry, University of the Basque Country UPV/EHU, Apartado 644, E-48080 Bilbao, Spain}
\affiliation{IKERBASQUE, Basque Foundation for Science, Plaza Euskadi 5, 48009 Bilbao, Spain}
\begin{abstract}
In this work, we propose how to load and manipulate chiral states in a Josephson junction ring in the so called transmon regimen. We characterise these states by their symmetry properties under time reversal and parity transformations. We describe an explicit protocol to load and detect the states within a realistic set of circuit parameters and show simulations that reveal the chiral nature. Finally, we explore the utility of these states in quantum technological nonreciprocal devices. 
\end{abstract}

\maketitle

\section{Introduction}

Within the great variety of quantum technologies under development, mesoscopic superconducting circuits have emerged offering a rich breeding ground to test theoretical proposals and study new physics \cite{Introduction to QEM circuits,Microwave Photonics 2017}. 

The Josephson effect plays a fundamental role in these circuits since Josephson junctions (JJs) are naturally non-dissipative and nonlinear elements \cite{Frontiers of the Josephson effect}. Due to the improvement of quantum technologies in superconducting circuits, particular stable configurations in such Josephson arrays \cite{Fazio 2001,Fazio 2012} and its dynamics have been possible to study.

A recent line of research within quantum simulation is the emulation of synthetic gauge fields \cite{Synthetic gauge fields,Chiral ground state currents,Gauge potentials} which overcomes several difficulties that bring the direct application of real magnetic fields in superconducting circuits. 

The presence of a magnetic field, real or synthetic, implies that the system breaks discrete symmetries such as parity (P) or time-reversal (TR). We may say a system has TRS if evolving forward in time, and then reversing the evolution for the same amount of time, the system ends up at the initial state.

In this context, reciprocity is defined as the invariance of a system under the exchange of source and observer \cite{Electromagnetic nonreciprocity,A. Kamal thesis}. Thus, breaking TRS in a controlled way and obtaining a nonreciprocal response is crucial in the design of quantum communication gadgets. For instance, breaking reciprocal symmetry in Josephson junctions can give rise to superconducting devices such as isolators, gyrators, circulators, directional amplifiers and wave mixers. 

Isolators and circulators are necessary elements in most superconducting circuit experiments to shield the circuit from external noise sources and to extract the signals out of the circuit. Magnetic nonreciprocal devices based on the Faraday effect involve centimetres sized magnets hindering the scalability of the circuits \cite{Microwave gyrator}. Nonetheless, there have been recent developments in the search of scalable, low noise, wide bandwidth, dynamical range nonreciprocal devices working at cryogenic temperatures \cite{A. Kamal 2011}. They serve the purpose of qubit readout in quantum computation, quantum simulation and quantum sensing and provide us with new capabilities. These proposals include a graph-based scheme to optimise nonreciprocal circuits \cite{Graph based analysis}, quantum Hall effect based gyrators and circulators \cite{Hall effect circulator,Quantum Hall circulator,Self impedance circulator}, parametric and traveling-wave parametric amplifiers \cite{Byeong 2012,near_quantum_limited_TWPA,Flux driven JPA,Low noise kinetic inductance,Optimizing Josephson ring modulator,Widely tunable parametric amplifier,M. H. Devoret 2010,Nonlinearities and parametric amp}, Josephson parametric converters \cite{A. Kamal 2013}, an interferometric Josephson isolator \cite{Baleegh Abdo 2019}, a field programmable Josephson amplifier \cite{Lecoc 2017}, mechanical circulators \cite{Mechanical on-chip microwave}, reconfigurable circulators \cite{Nonreciprocal reconfigurable circuit,Reconfigurable Josephson circulator}, a passive circulator \cite{Passive on-chip}, or others \cite{Kerckhoff 2015}.

In this article, we study chiral states \cite{ChiralSpinStates} in a Josephson junction ring \cite{Quantum phase slips} as natural quantum states that break P and TR symmetries. We show that breaking time-reversal symmetry can be achieved with a Josephson ring, in the so-called transmon regime, coupled to input/output ports also through JJs. In a quenched dynamics simulation, we calculate the lifetime of these states and characterise the out-of-equilibrium properties of this setup. Inspired by Koch {\it et al.} \cite{Time reversal} and M\"uller {\it et al.} \cite{Mueller_2018}, we show that a circulating behaviour is realised changing to a basis of common and differential input modes. Furthermore, a tunable directional coupler is proposed using the nonreciprocal features of the scattering matrix of three transmission lines connected to a JJ ring.  

\section{Chirality}
\label{sec: Chirality}

A superconducting node is described locally by a periodic degree of freedom $\{ | \phi \rangle \}$, where $\phi \in ( -\pi , \pi ]$ characterises the superconducting phase of a given superconducting island. Equivalently, the discrete conjugate variable $\{ | \tilde{n} \rangle \}$, where $\tilde{n} \in \mathbb{Z}$, characterises the number of Cooper pairs in the same superconducting island. This two local basis are related by a Fourier transform: $|\phi \rangle = \sum_{\tilde{n} \in \mathbb{Z}} e^{-i\phi \tilde{n}} |\tilde{n}\rangle$.

In a triangular plaquette given by three superconducting nodes connected by three Josephson junctions, a complete basis for the three superconducting nodes is given by $\{ | \phi_{1} \rangle \otimes | \phi_{2} \rangle \otimes | \phi_{3} \rangle \}$, in the flux basis, or $\{ | \tilde{n}_{1} \rangle \otimes | \tilde{n}_{2} \rangle \otimes | \tilde{n}_{3} \rangle \}$, in the charge basis. Again, both bases are related by a Fourier transform: $ | \phi_{1} ,  \phi_{2} , \phi_{3} \rangle = \sum_{\{\tilde{n}_{1},\tilde{n}_{2},\tilde{n}_{3}\} \in \mathbb{Z}} e^{-i\phi_{1} \tilde{n}_{1}} e^{-i\phi_{2} \tilde{n}_{2}} e^{-i\phi_{3} \tilde{n}_{3}} |\tilde{n}_{1} , \tilde{n}_{2} , \tilde{n}_{3} \rangle$.

If we consider the set of states with a fixed total charge $N=\tilde{n}_{1} + \tilde{n}_{2} + \tilde{n}_{3}$ in the flux basis, it can be described by
\begin{equation*}
|N,\varphi_{2},\varphi_{3} \rangle =\sum_{\{n_{2},n_{3}\} \in \mathbb{Z}} e^{-i\varphi_{2} n_{2}} e^{-i\varphi_{3} n_{3}} |N - n_{2} - n_{3} , n_{2} , n_{3} \rangle
\end{equation*}

From this set, we would like to characterise the subset of states that are invariant under the cyclic permutation of the three nodes $P_{123}$ right-handed (or $P_{132}$ left-handed), 
\begin{equation*}
\begin{split}
P_{123}|N,\varphi_{2},\varphi_{3} \rangle &= \sum e^{-i\varphi_{2} n_{2}} e^{-i\varphi_{3} n_{3}} | n_{2} , n_{3}, N - n_{2} - n_{3} \rangle \\
P_{132}|N,\varphi_{2},\varphi_{3} \rangle &= \sum e^{-i\varphi_{2} n_{2}} e^{-i\varphi_{3} n_{3}} | n_{3} , N - n_{2} - n_{3} , n_{2} \rangle 
\end{split}
\end{equation*}

These states are equivalent up to an overall phase $|N,\varphi_{2},\varphi_{3} \rangle \sim P_{123}|N,\varphi_{2},\varphi_{3} \rangle \sim P_{132}|N,\varphi_{2},\varphi_{3} \rangle$ if $3 \varphi_{2} =2 \pi \mathbb{Z}$, $3 \varphi_{3} =2 \pi \mathbb{Z}$, and $\left( \varphi_{2} + \varphi_{3} \right)=2 \pi \mathbb{Z}$ which give us just three states:
\begin{equation*}
\begin{split}
|N,0,0 \rangle &=\sum  |N - n_{2} - n_{3} , n_{2} , n_{3} \rangle \\
|N,\frac{2 \pi}{3},-\frac{2 \pi}{3} \rangle &=\sum e^{-i \frac{2 \pi  }{3} n_{2}} e^{ i\frac{ 2 \pi }{3} n_{3} } |N - n_{2} - n_{3} , n_{2} , n_{3} \rangle \\
|N,-\frac{2 \pi}{3},\frac{2 \pi}{3} \rangle &=\sum e^{ i\frac{ 2 \pi }{3} n_{2}} e^{-i \frac{ 2 \pi }{3} n_{3} } |N - n_{2} - n_{3} , n_{2} , n_{3} \rangle 
\end{split}
\end{equation*}

It is straightforward to realise that under the action of time-reversal or parity transformation, $|N,0,0 \rangle$ remains invariant, while $|N,\frac{2 \pi}{3},-\frac{2 \pi}{3} \rangle$ maps onto $|N,-\frac{2 \pi}{3},\frac{2 \pi}{3} \rangle$.

The two non-trivial states under the action of these permutations are the only two chiral states in this setup. In fact, it can be defined a chiral operator \cite{ChiralSpinStates} $\chi = \frac{P_{123} - P_{132}}{2i}$ which changes the sign under $P$ or  $TR$ transformation but remains invariant under the combination of parity and time reversal $(PT)$ transformation. In this way, a non-zero value of this operator signals $P$ and $T$ symmetry breaking states. 

More concretely, the eigenvalues of the permutation operator for these three states are:  $P_{123}\big|_{ |N,0,0 \rangle} =1$, $P_{123}\big|_{ |N,\frac{2 \pi}{3},-\frac{2 \pi}{3} \rangle} =e^{-i \frac{2 \pi  }{3} N}$, and $P_{123}\big|_{ |N,-\frac{2 \pi}{3},\frac{2 \pi}{3} \rangle }=e^{i \frac{2 \pi  }{3} N}$, which gives $\chi |N,\frac{2 \pi}{3},-\frac{2 \pi}{3} \rangle =- \sin{\left( \frac{2 \pi N}{3} \right)}  |N,\frac{2 \pi}{3},-\frac{2 \pi}{3} \rangle$, $\chi |N,-\frac{2 \pi}{3},\frac{2 \pi}{3} \rangle = \sin{\left( \frac{2 \pi N}{3} \right)}  |N,-\frac{2 \pi}{3},\frac{2 \pi}{3} \rangle$, and $\chi  |N,0,0 \rangle = 0$. In fact, the phase acquired by the state under the permutation $P_{123}$ can be understood as a Berry phase. 

\begin{figure}[!]
\centering
\includegraphics[width=0.75\linewidth]{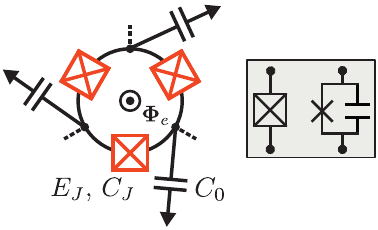}
\caption{ Circuit representation of the triangular plaquette given by three superconducting nodes connected by three JJs threaded by an external magnetic flux $\Phi_{e}$ with capacitive bias to ground; for the measurement and characterisation of the triangular plaquette, three transmission lines are connected with three external JJs and three resonators. The main parameters of the circuit that will be used through the text are: $C_{0}$, the capacitance to the ground of each node, $E_{J}$, $C_{J}$ are the Josephson energy and capacitance of the JJs ring respectively.}
\label{fig1}
\end{figure}

\section{Circuit QED architecture}

Following \cite{Time reversal,Mueller_2018}, we consider a minimal circuit QED (cQED) set up consisting of a ring of three Josephson junctions threaded by an external magnetic flux with capacitive bias to ground (see Fig. \ref{fig1}). By making use of flux-node description \cite{Devoret_1995_QFluct}, the Lagrangian of the system can be directly written 
\begin{eqnarray}
	L&=& \frac{1}{2}\dot{\bsb{\Phi}}^T\msf{C}\dot{\bsb{\Phi}}+\sum_i E_{Ji}\cos(\Delta\phi_i-\phi_{e,i}),
\end{eqnarray}
where the vector of node fluxes is defined $\bsb{\Phi}\equiv(\Phi_1, \Phi_2, \Phi_3)$, the phase variables are defined through the second Josephson relation $\phi_x=2\pi\Phi_x/\Phi_0$ with $\Phi_0$ the flux quantum constant. The differences of phases correspond to $\Delta\phi_i=\phi_{i+1}-\phi_i$, for $i=\{1,2,3\}$ and identifying $i=4$ with $i=1$. The capacitance matrix is defined as 
\begin{equation}
	\msf{C}=\begin{pmatrix}
		C_{\Sigma}&-C_J&-C_J\\
		-C_J&C_{\Sigma}&-C_J\\
		-C_J&-C_J&C_{\Sigma}\\	
\end{pmatrix},
\end{equation}
with $C_{\Sigma}=C_0+2C_J$, and $C_{0}$ local capacitance and $C_{J}$ the capacitance in the Josephson junctions. The external flux $\phi_{e}$ appears naturally equally distributed among the three links, i.e., $\phi_{e,i}=\phi_{e}/3$, with this gauge choice, the translational invariance is not explicitly broken.

The Legendre transformation involves the definition of conjugated charge variables $\bsb{Q}=\msf{C}\dot{\bsb{\Phi}}$, to derive the Hamiltonian
\begin{eqnarray}
	H&=& \frac{1}{2}\bsb{Q}^T \msf{C}^{-1}\bsb{Q}  -\sum_i E_{Ji}\cos(\Delta\phi_i-\frac{\phi_{e}}{3}).\label{eq:H_plaquette_ext_flux}
\end{eqnarray}
For the sake of simplicity, let us work with the number of Cooper pair variables $\bsb{\tilde{n}}=\bsb{Q}/2e$, where $e$ is the electron charge. Conjugated classical variables are promoted to operators, with commutation relations $[\tilde{n}_i,  e^{\mp i\phi_j}]=\mp\delta_{ij}e^{\mp i\phi_j}$.

We recall that Hamiltonian (\ref{eq:H_plaquette_ext_flux}) has an important symmetry, readily, the total charge in the plaquette $N=\sum_i \tilde{n}_i$ is a conserved quantity. We can perform a canonical transformation $\bsb{\varphi}\rightarrow\msf{T}\bsb{\varphi}$ and $\bsb{n}\rightarrow(\msf{T}^T)^{-1}\bsb{n}$, 
\begin{equation}
	\msf{T}=\begin{pmatrix}
		1 & 0 & 0 \\
		-1 & 1 & 0 \\
		-1 & 0 & 1 \\
	\end{pmatrix}
	\label{eq: first canonical transformation}
\end{equation}
which defines $\varphi \equiv \phi_{1}$, $\varphi_{2} \equiv \phi_{2} - \phi_{1}$, $\varphi_{3} \equiv \phi_{3} - \phi_{1}$ and the conjugate momenta or charge operators $N \equiv \tilde{n}_{1}+\tilde{n}_{2}+\tilde{n}_{3}$, $n_{2} \equiv \tilde{n}_{2}$, $n_{3} \equiv \tilde{n}_{3}$ from which we arrive at the Hamiltonian
\begin{equation}
\begin{split}
&H=E_{N} N^{2} - E_{J} ~ V \left(\varphi_{2} , \varphi_{3} \right) \\
&+E_{C} \left[ \left( N-n_{2}-n_{3} \right)^{2} + n_{2}^{2} + n_{3}^{2} \right]. 
\label{eq: Plaquette Hamiltonian}
\end{split}
\end{equation}
with $E_{N}= \frac{2e^{2}C_{J}}{C_{0} \left(C_{0} + 3 C_{J} \right) } $, $E_{C}= \frac{2e^{2}}{\left(C_{0} + 3 C_{J} \right)}$, $V \left(\varphi_{2} , \varphi_{3} \right)= \cos{\left( \varphi_{2} - \frac{\phi_{e}}{3} \right)} + \cos{\left( \varphi_{3} + \frac{\phi_{e}}{3} \right)} + \cos{\left( \varphi_{3} - \varphi_{2} - \frac{\phi_{e}}{3} \right)}$. For possible sources of disorder in the dynamics or decay channels see Appendix \ref{disorderanddecay}

\begin{figure}[!]
\centering
\includegraphics[width=1\linewidth]{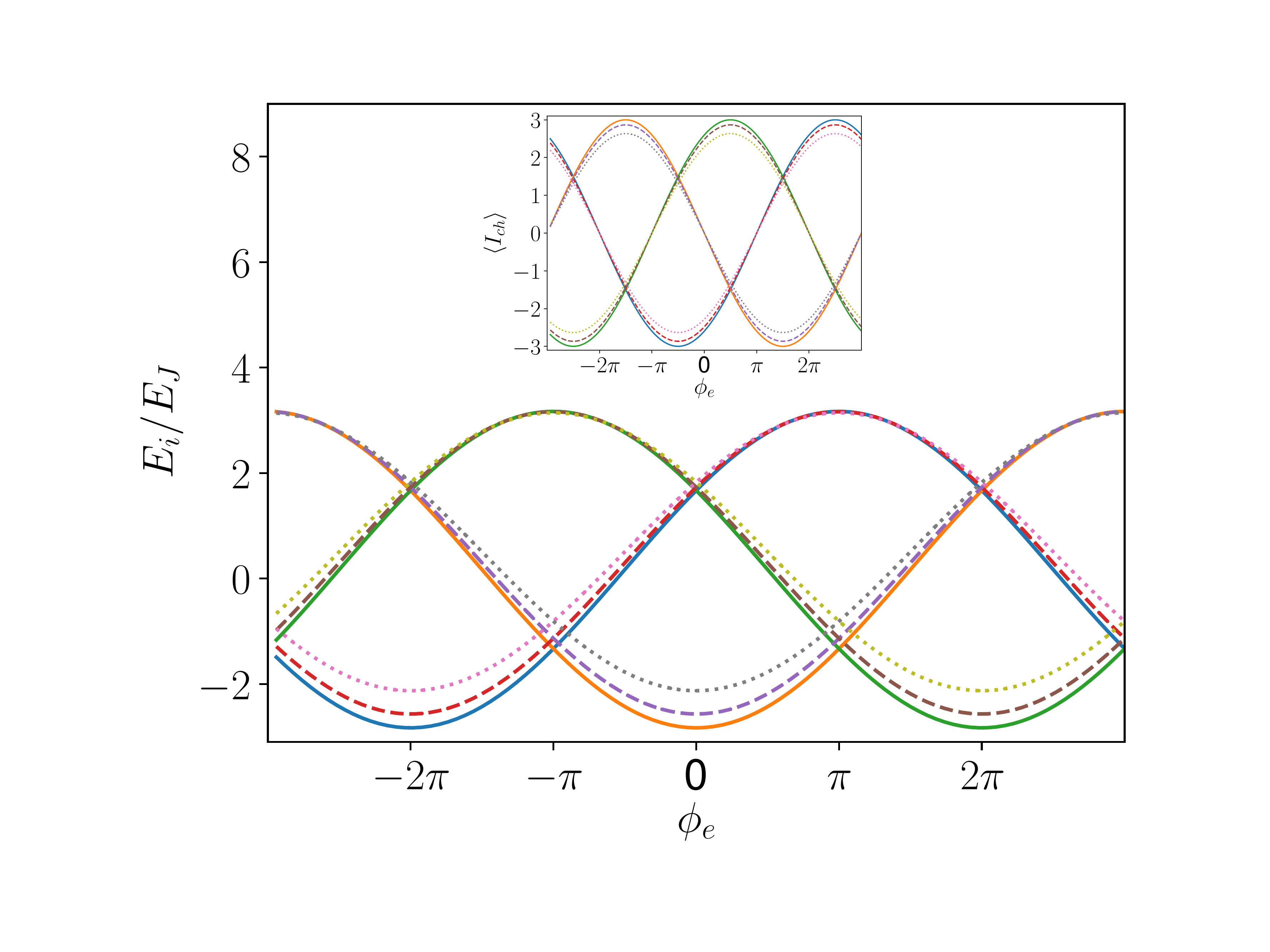}
\caption{Spectrum of Hamiltonian \eqref{eq: Plaquette Hamiltonian} and its dependence on the external flux. The inset shows this dependence for the expected value of the chiral current. The continuous lines refer to the energies and currents for a ratio $E_{J}/E_{C} = 10^{5}$, the dashed lines for $E_{J}/E_{C} = 10^{2}$ and the dotted ones to  $E_{J}/E_{C} = 10$, being $E_{J} = 10 GHz$ and considering one excitation $N=1$. When the external flux $\phi_{e} \in \left[ -3 \pi, 3 \pi \right]$ there are three special ground states, one centred at $-2 \pi$ for the plaquette being in the state $\ket{N,-\frac{2\pi}{3},\frac{2\pi}{3}}$, another for a zero flux which corresponds to the state $\ket{N,0,0}$ and the one for a $2 \pi$ flux and the state $\ket{N,\frac{2\pi}{3},-\frac{2\pi}{3}}$. It is important to note that the spectrum of the Hamiltonian maps to itself whenever we introduce a transformation $\phi_{e} \rightarrow \phi_{e} +  2 \pi k$ with $k \in \mathbb{Z}$. We take advantage of this spectral flow in order to load one of the two chiral states in the plaquette.}
\label{fig: spectral flow}
\end{figure}

In the following, we will use this Hamiltonian in the limit $E_{N} \gg  E_{J} \gg E_{C}$ (or $\frac{e^{2}}{C_{0}} \gg E_{J} \gg \frac{e^{2}}{C_{J}}$). Therefore, neglecting the last term in the previous Hamiltonian, the eigenvectors are given by the vectors $|N,\varphi_{2},\varphi_{3} \rangle$ just defined in the later section, and as a function of the external flux $\phi_{e}$, the ground state of this Hamiltonian is given by: $|N,0,0 \rangle$ when $|\phi_{e}| < \pi$; $ |N,\frac{2 \pi}{3},-\frac{2 \pi}{3} \rangle$ when $\pi < \phi_{e} < 3 \pi$; and $ |N,-\frac{2 \pi}{3},\frac{2 \pi}{3} \rangle$ when $-3 \pi < \phi_{e} < - \pi$ (see Fig. \ref{fig: spectral flow}).

The spectrum of the Hamiltonian does not change when $\phi_{e}$ is changed by $2\pi \mathbb{Z}$, nonetheless the eigenvectors do not remain the same in the flow of changing the external flux. This fact characterises the spectral flow of the Hamiltonian that we will use to load the chiral states in the ring.

Another figure of merit we will use to characterise the states loaded in the triangular plaquette is given by the chiral current, that is just the sum of the current at every Josephson junction, i.e. $I_{\text{ch}} (\phi_{e}) =I_{0} \sum_i \sin(\Delta\phi_i-\phi_{e}/3) =I_{0}  \sin{\left( \varphi_{2} - \frac{\phi_{e}}{3} \right)} - I_{0} \sin{\left( \varphi_{3} + \frac{\phi_{e}}{3} \right)} + I_{0} \sin{\left( \varphi_{3} - \varphi_{2} - \frac{\phi_{e}}{3} \right)}$.  At $\phi_{e} =0$, $I_{\text{ch}} (0) $ changes the sign under $P$ or $TR$ transformation, such that a non-zero expectation value of this operator can signal $P$ and $T$ symmetry breaking states. 

\begin{figure}[!]
\includegraphics[scale=0.23]{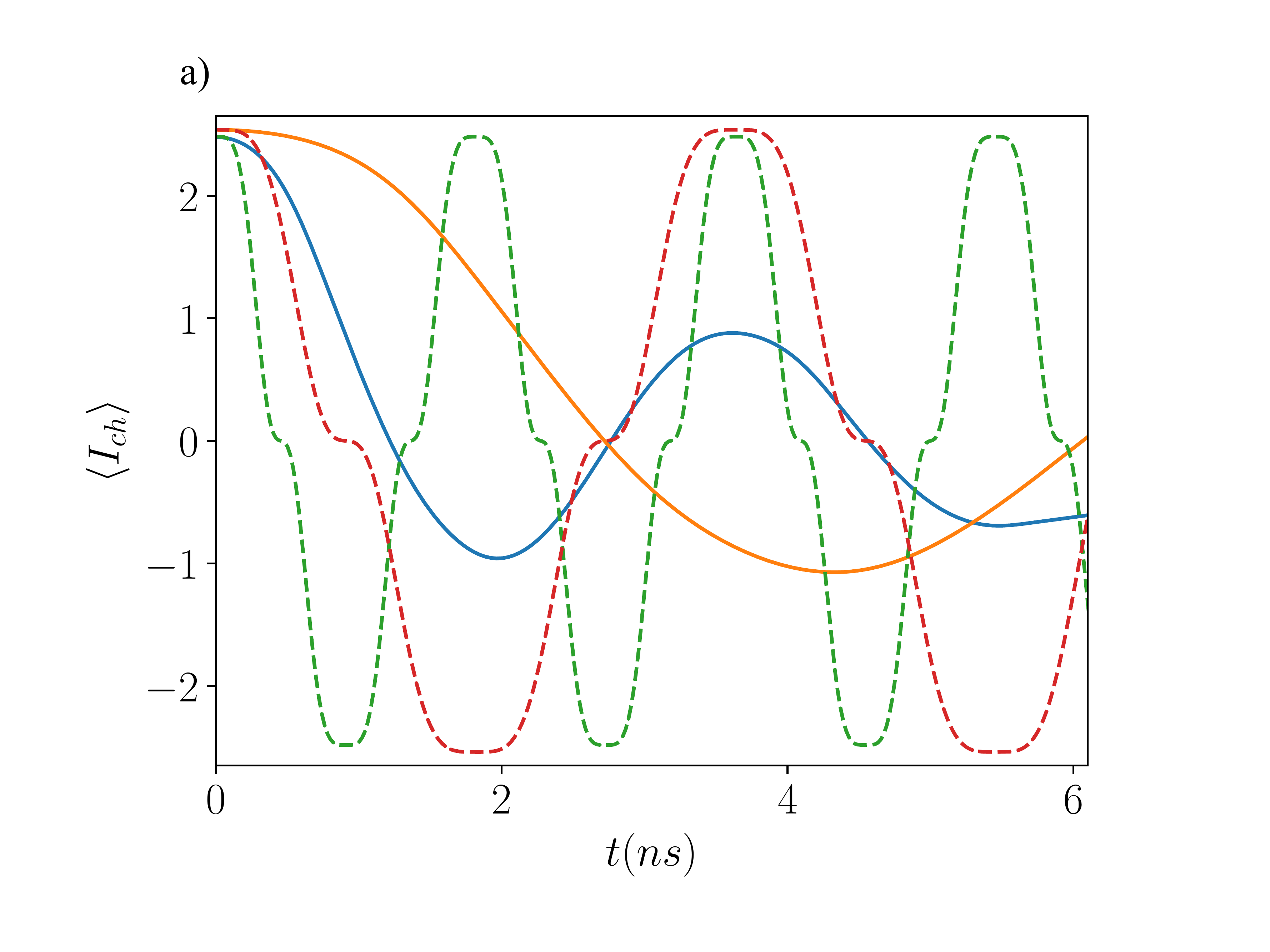}
\includegraphics[scale=0.23]{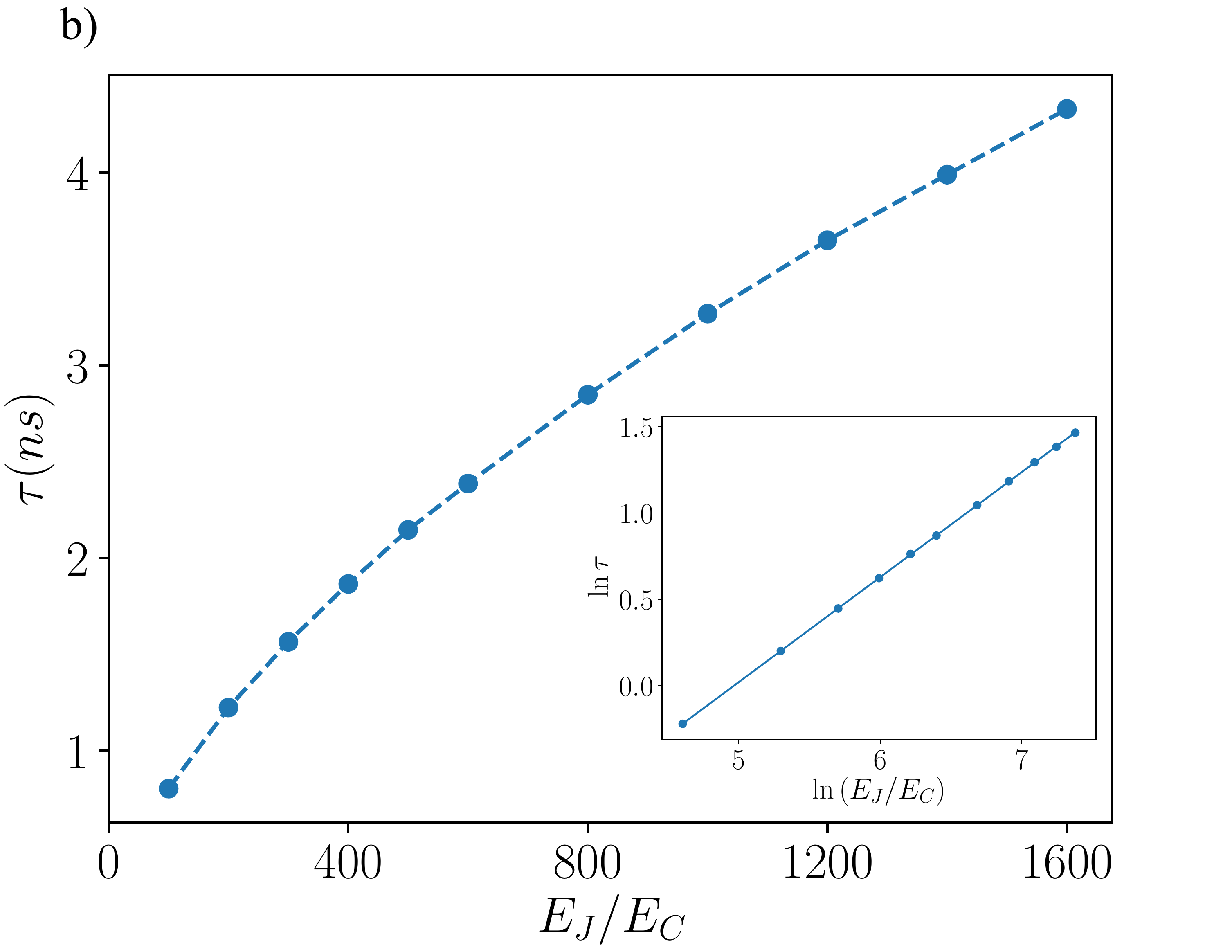}
\caption{(a) Chiral current expected value time evolution obtained for two different $E_J/E_C$ ratios. The dashed plots correspond to the currents calculated within the harmonic approximation. The blue and green plots stand for $E_J/E_C = 10^{2}$ and the orange and red curves for $E_J/E_C = 4 \times 10^{2}$. The higher the ratio, the bigger the number of points we need to use in the discretisation to reach the continuum limit (see Appendix \ref{sec: numeric estimation}). (b) Half-life time dependence on the $E_J/E_C$ ratio numerical values. In this plot, we have investigated higher ratios with the exact Hamiltonian in the phase basis. The half-life time is the time it takes the chiral current to halve its initial value $ \langle I_{ch} \rangle \left(t = \tau \right) = \langle I_{ch}\rangle \parent{t=0} /2 $. From the numerical fit of the curve $\frac{\tau}{\tau_{0}} = \left( \frac{E_{J}}{E_{C}} \right)^{\alpha}$, we extract $\alpha=0.6088601 \pm 6 \times 10^{-7}$ and $\tau _{0} = 0.04859 \pm 2 \times10^{-5} \mathrm{ ns}$. The $\tau$ dependence on the $E_J/E_C$ ratio predicts that a $100 \mathrm{ns}$ half-life time  can be achieved for $E_{J}/E_{C}$ of the order of $10^{5}$.}
\label{fig: chiral currents}
\end{figure}

\subsection{Spectral flow}

A Josephson ring can be described with the Hamiltonian (\ref{eq:H_plaquette_ext_flux}), where the classical magnetic flux $\phi_{e}$ is shared equally by every junction and thus translational invariance remains an explicit symmetry. 

This Hamiltonian is unitarily equivalent and therefore iso-spectral to another Hamiltonian $H_{1}$ where the potential energy is given by $ V_{1} \left(\varphi_{2} , \varphi_{3} \right)= \cos{\left( \varphi_{2}  \right)} + \cos{\left( \varphi_{3}  \right)} + \cos{\left( \varphi_{3} - \varphi_{2} - \phi_{e} \right)}$ where the magnetic flux just appears on one of the JJs. The unitary transformation that maps $H$ onto $H_{1}$ is composed by a sequence of phase displacements. For instance, starting with a displacement of $\varphi_{3} \to \varphi_{3}-\frac{\phi_{e}}{3}$, followed by $\varphi_{2} \to   \varphi_{2}+\frac{2\phi_{e}}{3}$ we recover $H_{1}$.

A particular and clarifying limit is given by the classical magnetic flux $\phi_{e} = 2 \pi k$ with $k \in \mathbb{Z}$. At this value, $V \left( \varphi_{2} , \varphi_{3} \right) = \cos{\left( \varphi_{2} - \frac{2 \pi k}{3} \right)} + \cos{\left( \varphi_{3} + \frac{2 \pi k}{3} \right)} + \cos{\left( \varphi_{3} - \varphi_{2} - \frac{2 \pi k}{3} \right)}$ and $V_{1} \left( \varphi_{2} , \varphi_{3} \right) = \cos{\left( \varphi_{2}  \right)} + \cos{\left( \varphi_{3}  \right)} + \cos{\left( \varphi_{3} - \varphi_{2}  \right)}$.

We can be tempted to assume as ``trivial'' the action of the magnetic flux at any of these $k$ points. Nonetheless, $H$ and $H_{1}$ are just iso-spectral but there is a non-trivial unitary action on the eigenstates. Under the displacement operator $e^{- \frac{ i n_{3} 2 \pi }{3}} e^{  \frac{ i n_{2} 2 \pi }{3}}$, the states $|N,\varphi_{2},\varphi_{3} \rangle \to |N,\varphi_{2} + \frac{2 \pi}{3},\varphi_{3} - \frac{2 \pi}{3}\rangle $, and in particular $|N,0,0\rangle \to |N, \frac{2 \pi}{3} , - \frac{2 \pi}{3} \rangle \to |N, -\frac{2 \pi}{3} , \frac{2 \pi}{3} \rangle \to |N, 0 , 0 \rangle$.

\section{Superconducting chiral states}
\label{sec: preparation of the states}
In the following, we describe how to load and detect the chiral states. These states are characterised by the appearance of currents flowing clockwise or counter-clockwise through the loop. 

The first step to prepare the initial state is to thread the plaquette with a magnetic flux of $2 \pi$, in units of the magnetic flux quantum $\Phi_{0}$. After that, the system is cooled down until it reaches the ground state shown in Fig. \ref{fig: spectral flow}, which is a chiral state. Then, we perform a sudden quench by turning off the magnetic flux. In this way, the chiral state becomes a highly excited state of the free Hamiltonian. 

When an external static magnetic flux $\phi_{e}$ threads the ring, the Hamiltonian is given by equation \eqref{eq: Plaquette Hamiltonian}. We work in the phase regime in which Josephson energy is much bigger than the charge energy $E_{N} \gg E_{J} \gg E_{C}$. Being the total charge a conserved quantity ($N$ is a good quantum number) we can restrict ourselves to a subspace of constant total number of excitations. Applying the canonical transformation
\begin{equation}
\begin{split}
\phi _+ = \frac{1}{2} \parent{\varphi _2 + \varphi _3},& \quad \phi _- = \frac{1}{2} \parent{\varphi _2 - \varphi _3},\\
n_+ = n_2 + n_3, &\quad n_- = n_2-n_3,
\end{split}
\end{equation}
such that the new variables satisfy the usual commutation relations $\left[n_{\pm}, e^{i \phi _{\pm}} \right] = e^{i \phi _{\pm}}$, the Hamiltonian is mapped onto
\begin{equation*}
\begin{split}
&H  = \left(E_{N} + \frac{E_{C}}{3}\right) N^2 + \frac{E_C}{2} \cor{3 \parent{n_+ - \frac{2}{3}N }^2 + n_-^{2}} \\
&-E_J\cor{2 \cos \phi _+ \cos \parent{\phi _- - \frac{\phi_{e}}{3}}  + \cos \parent{2 \phi _- + \frac{\phi_{e}}{3}}}.
\end{split}
\end{equation*}
To gain more insight we also study the Hamiltonian in the harmonic approximation around $\phi_{+} \to 0$ and $\phi_{-} \to \frac{\phi_{e}}{3} = \frac{2 \pi k}{3}$, $k \in \mathbb{Z}$ , where
\begin{equation}
\begin{split}
H &\to \left(E_{N} + \frac{E_{C}}{3}\right)N^2 + \frac{E_C}{2} \cor{3 \parent{n_+ - \frac{2}{3}N }^2 + n_-^{2}} \\
& + E_{J} \cor{\phi _{+}^{2} + 3 \parent{ \phi _{-}- \frac{\phi_{e}}{3}}^{2}}.
\end{split}
\label{eq: harmonic Hamiltonian}
\end{equation}
To perform the numerical calculations, we discretise the phase degrees of freedom. The phases are chosen to take $L$ values $\phi _{\pm} \equiv \frac{2 \pi k_{\pm}}{L}$ contained in the interval  $\phi _{\pm} \in \left(-\pi ,\pi \right]$ (or $k_{\pm} \in \left[ - \frac{L}{2}+1, \frac{L}{2} \right]$), setting the charges to lie in $n_{\pm} \in \left[ - \frac{L}{2}+1, \frac{L}{2} \right]$. In the limit of $L \rightarrow \infty $ the continuum is recovered. 

The representation of the ground state of the Hamiltonian requires a minimum number of discrete levels $L$ to achieve a faithful numerical simulation, which indeed depends on the ratio $E_{J}/E_{C}$ (see Appendix \ref{sec: numeric estimation}). Moreover, we evolve the chiral current operator with an increasing number of discrete levels $L$ till the curves of the evolution collapse to the continuum limit. The higher the energy ratio $E_{J}/E_{C}$ the bigger number of levels are needed to reach the continuum limit. 

We are interested in the regime where the net chiral current flowing in one sense is nonzero, so that the state breaks TRS. For this reason, we define the time $\tau$ as the time it takes the current to halve its initial value, so that the chiral properties of the state are still manifested. The dynamics in the harmonic approximation can be completely described in terms of coherent states, and the chiral current oscillates as expected.
We take the chiral current operator as the sum of currents flowing through the three nodes of the plaquette, which in the variables we have chosen 
\begin{equation*}
I_{ch} = 2 \cos \phi _+ \sin \parent{\phi _- - \frac{\phi_{e}}{3}}- \sin \parent{2\phi _- +  \frac{\phi_{e}}{3}}.
\end{equation*}
Expressing the Hamiltonian in the phase basis we diagonalise it numerically to obtain the ground state for a fixed value of the magnetic flux $\Phi_{e}=2\pi \Phi_{0}$. Next, we evolve the chiral state in time with the Hamiltonian without magnetic flux and we compute the time evolution of the chiral current. We find the chiral current lives longer according to a power law on the ratio $E_{J}/E_{C}$. It is important to mention that the state of the plaquette is specially robust against charge noise, as we are working in the phase regime. Once we have obtained the chiral current, we seek for the regime in which it preserves a single circulation sense and take the half-life time to characterise this regime. 

In Fig. \ref{fig: chiral currents}, we fit the numerical data for the mean lifetime of the chiral current to the curve $\frac{\tau}{\tau_{0}} = \left( \frac{E_{J}}{E_{C}}\right)^{\alpha}$ with the numerical parameters $\alpha=0.6088601 \pm 6 \times 10^{-7}$ and $\tau _{0} \sim 0.04859 \pm 2  \times 10^{-5} \mathrm{ ns}$. Therefore, currents with $\tau$ in the order of a hundred nanoseconds may be accomplished for $\frac{E_{J}}{E_{C}} \approx 10^{5}$.

\subsection{Chiral effective dynamics}

In order to introduce a weak perturbation in the JJ ring to measure the state of the ring, we couple a resonator of frequency $\omega _r \ll E_J$ to each of the three nodes in the plaquette such that the energy levels of the JJ ring are much closer among them than those of the resonators. With this in mind we can decouple and eliminate the degrees of freedom of the ring and derive a low-energy Hamiltonian for the resonators. As the coupling elements we choose are JJs, the total number of charges in the ring is no longer preserved. If we remain in the one excitation subspace of the plaquette, this extra charge will be able to hop to the adjacent resonators while the chiral state remains. For the calculation of the effective Hamiltonian and input-output relations of the next subsection, we follow \cite{Time reversal}. Taking the limit of $E_{C} \ll E_{J}$ in the Hamiltonian of the plaquette 
\begin{figure}[!]
\begin{minipage}[h]{\linewidth}
\includegraphics[width=.8\linewidth]{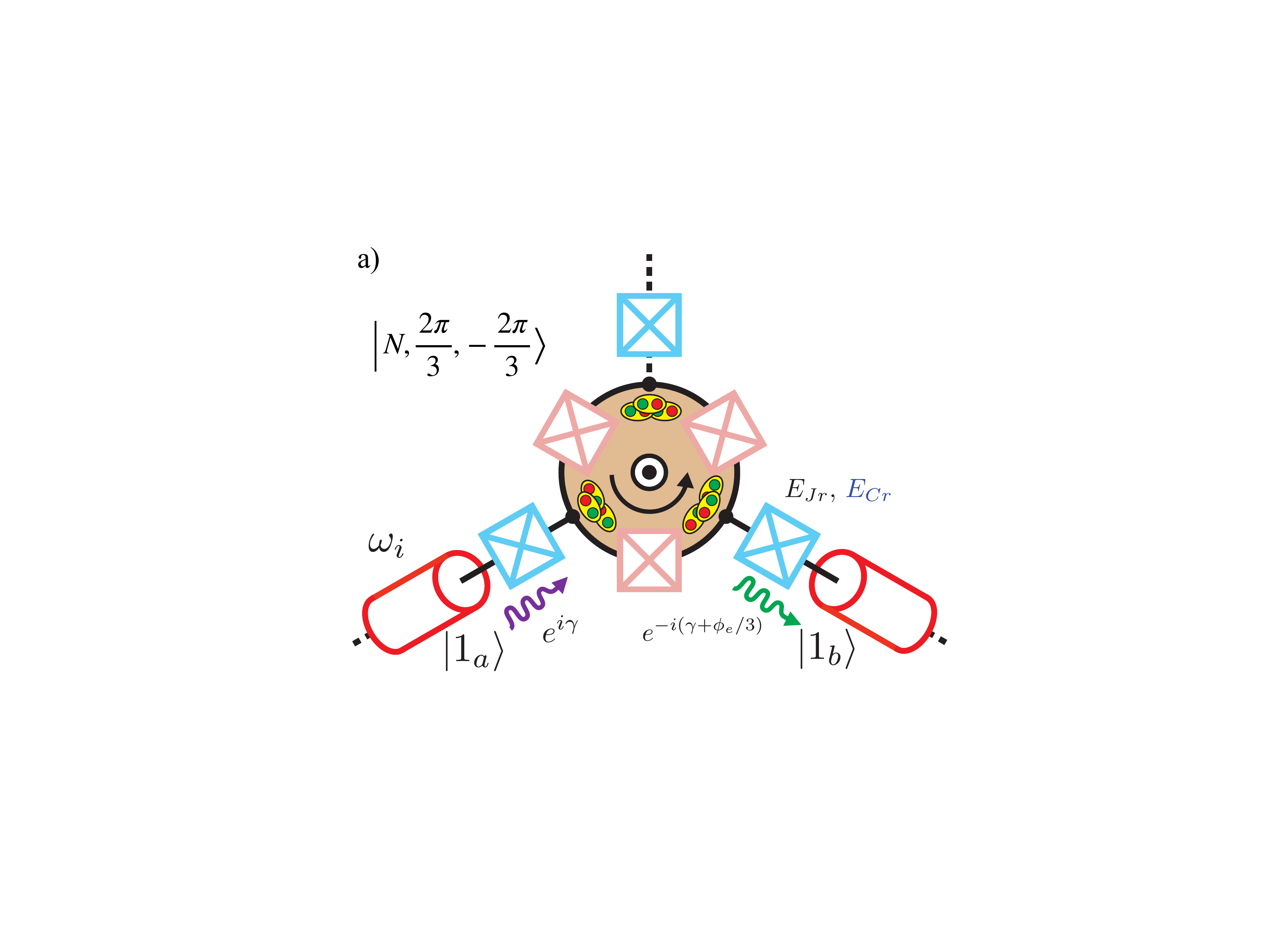}
\end{minipage}\vfill%
\begin{minipage}[h]{\linewidth}
\includegraphics[width=.8\linewidth]{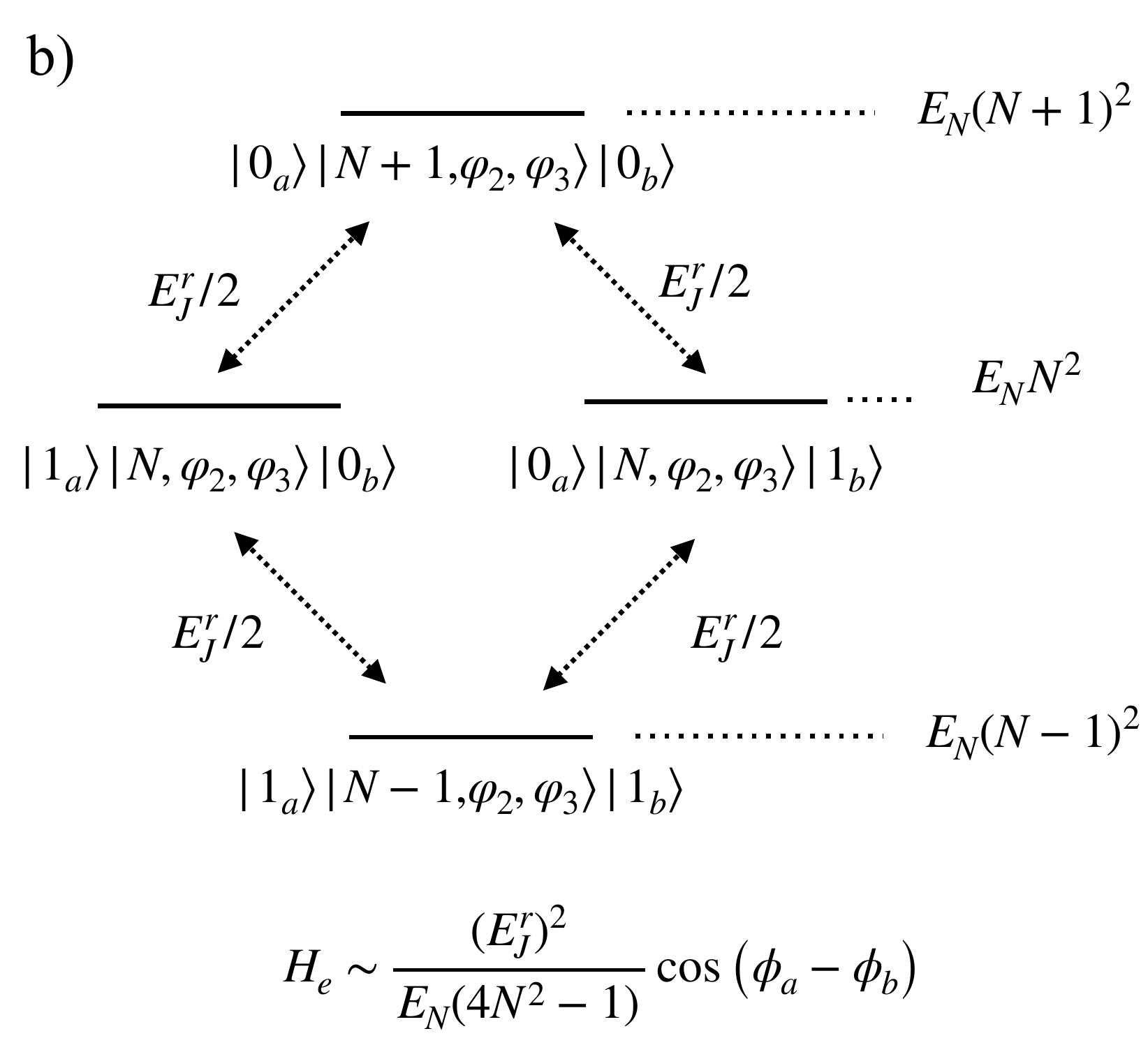}
\end{minipage}\vfill%
\begin{minipage}[h]{\linewidth}
\includegraphics[width=.8\linewidth]{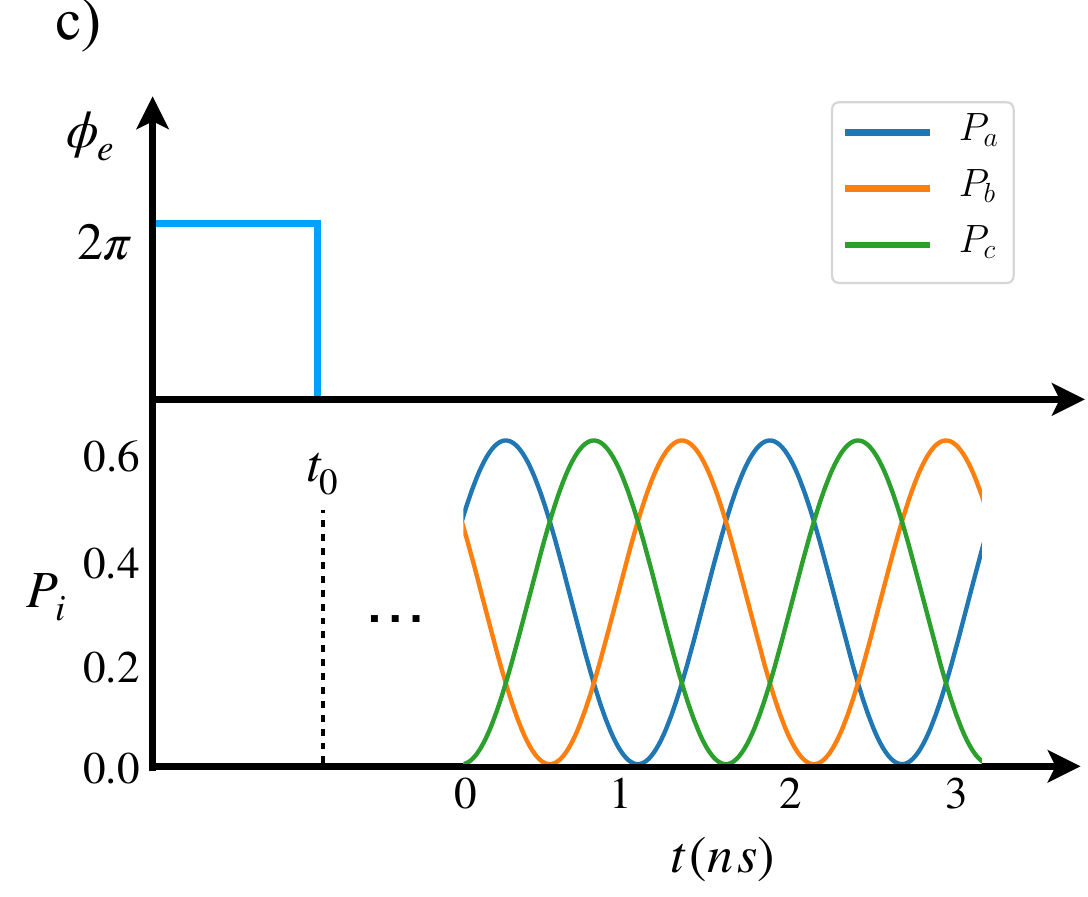}
\end{minipage}
\caption{(a) Schematic representation of the hopping of one excitation from state $\ket{1_{a}}$ to the state $\ket{1_{b}}$. The excitation in the first resonator gains a phase $e^{i \gamma}$ when it tunnels through the JJ connecting the resonator to the ring. This phase is lost when it exits the ring towards the adjacent resonator such that the only remaining phase is the one it acquires in the ring. (b) Energy level description of the JJ triangular plaquette and the effect of the Josephson coupling to the three external nodes. (c) System quench and transition probabilities for one excitation to hop from resonator to resonator when the initial state of the resonators is $\frac{1}{\sqrt{2}} \parent{\ket{100} - \ket{010}}$. Once the ring has relaxed to its ground state the magnetic field is switched off, leaving the plaquette in the chiral state $\ket{N, \frac{2 \pi}{3}, - \frac{2 \pi}{3}}$. The state considered in the resonators is non-chiral, that is to say, the expected value of $\chi$ in this state is zero. Hence, the circulation in the resonators is a signature of the plaquette hosting a chiral state.}
\label{fig: circulation}
\end{figure}

\begin{equation}
\begin{split}
H = & E_{N} N^{2} - E_{J} \left[2 \cos \phi _+ \cos \parent{\phi _- - \frac{\phi_{e}}{3}} \right. \\
&\left. + \cos \parent{2 \phi _- + \frac{\phi_{e}}{3}} \right] + \sum _{i=a,b,c} \omega _{i} a^{\dagger}_{i} a_{i} +  H_{\text{int}},
\end{split}
\end{equation}
where $\omega _{i} = \omega _{r}$ are the frequencies of the resonators which we assume to be the same, and the interaction Hamiltonian 
\begin{equation*}
H_{\text{int}} = \frac{E_{J}^{r}}{2}  \parent{e^{i \parent{\phi _1 - \phi_a}}+e^{i \parent{\phi _2 - \phi _b}}+e^{i \parent{\phi _3 - \phi _c}} + \text{h.c.} }.
\label{eq: Interaction Hamiltonian}
\end{equation*}
$E_{J}^{r}$ is the Josephson energy of the JJs coupling the resonators with the ring, being the resonators $a,b$ and $c$ coupled to nodes $1,2$ and $3$ respectively. We consider the limit of $E_{C}^{r} \rightarrow 0$, \textit{i.e.} no kinetic energy for the resonators' junctions. This term would affect the phases of the state of the plaquette whereas $H_{\text{int}}$ leaves the state invariant and therefore does not destroy the chirality of the ring. The effective Hamiltonian is obtained through a Schrieffer-Wolf transformation \cite{Atom-Photon Interactions} and subsequent projection onto the chiral state of the plaquette

\begin{equation}
\begin{split}
H_{e} &= P_{plq} H P_{plq} + P_{plq} H_{\text{int}} P_{plq} \\
&+ \frac{1}{2} P_{plq}\cor{i E_J^r S , H_{\text{int}}} P_{plq} + \dots,
\end{split}
\end{equation}
with $S$ being the generator of the transformation and $P_{plq} = \ket{N, \frac{2 \pi}{3},-\frac{2 \pi}{3}}\bra{N, \frac{2 \pi}{3},-\frac{2 \pi}{3}}$ the projector for the plaquette in the chiral state. In the charge basis, the exponentials of the phases act as creation and annihilation operators such that the effective Hamiltonian 
\begin{equation}
\begin{split}
&H_{e} = \sum _{i=a,b,c} \parent{\omega _{i}+3g} a_{i}^{\dagger}a_{i} \\
&+ \frac{g}{2} \left( a_{b}^{\dagger} a_{a} e^{-i \frac{2\pi}{3}} + a_{c}^{\dagger} a_{b} e^{-i \frac{2\pi}{3}}  + a_{a}^{\dagger} a_{c} e^{-i \frac{2\pi}{3}} + \text{h.c.} \right),
\end{split}
\label{eq: effective Hamiltonian}
 \end{equation}

\begin{equation}
g = \frac{  \parent{E_{J}^r}^{2}}{E_{N} \parent{1- \parent{\frac{\omega _{r}}{E_{N} }-2N}^{2}}} \approx \frac{  \parent{E_{J}^r}^{2}}{E_{N }\parent{1-4N^{2}}},
\end{equation}
being in this case $N$ the constant denoting the number of excitations in the chiral state of the plaquette.

The phases that appear in the Hamiltonian are directly due to the initial state in which the triangular plaquette is loaded. A change in the external flux affects only the eigenvalues of the initial Hamiltonian of the plaquette but not the phases.  

The effective Hamiltonian can be diagonalised $H_{e} = \sum_{k} A^{\dagger}_{k}A_{k} \Omega _{k}$, the energies given by
\begin{equation}
\Omega _{k} = 3g +  \omega _{r}  + 2g \cos \parent{\frac{2 \pi k}{3}+ \frac{2\pi}{3}},
\label{eq: Omegas}
\end{equation}
where the subindex denotes the allowed wave-numbers $k=-1,0,1$ of the three eigenstates, as the Hamiltonian is diagonal in the reciprocal space. 

Consider now the case of introducing a single excitation in one resonator. We find the excitation can be observed subsequently in the other two resonators with equal likelihood, so neither circulation nor signature of chirality is observed in this scenario (see Appendix \ref{appendix: transition probabilities}).
If the initial state of the resonators is $\ket{\Psi _{0}} = \frac{1}{\sqrt{2}} \parent{\ket{100} - \ket{010}}$, where $\bra{\Psi _{0}} \chi \ket{\Psi _{0}} = 0$, which ensures we are not introducing any chirality by initiating the resonators in this state, the probability of finding the excitation in each resonator shows a clear circulating behaviour (see Fig. \ref{fig: circulation}).

\subsection{Nonreciprocal S-matrix}

\begin{figure}[!]
\centering
\includegraphics[width=\linewidth]{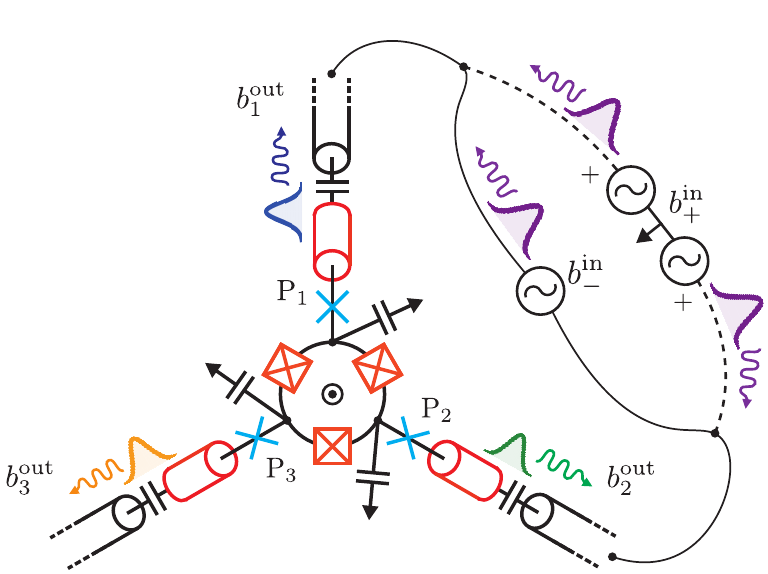}
\caption{Circuit representation of the triangular plaquette coupled to three resonators with frequencies $\omega_{i}$ through Josephson junctions with energies $E^{r}_{J}$ which are connected capacitively to transmission lines. Input mode $b_{+}^{in} = \frac{1}{\sqrt{2}}\parent{b_1^{in}+ b_2^{in}}$ and $b_{-}^{in} = \frac{1}{\sqrt{2}}\parent{b_1^{in}- b_2^{in}}$ are shown.}
\label{fig: circuit}
\end{figure}

\begin{figure}[!]
\includegraphics[scale=0.332]{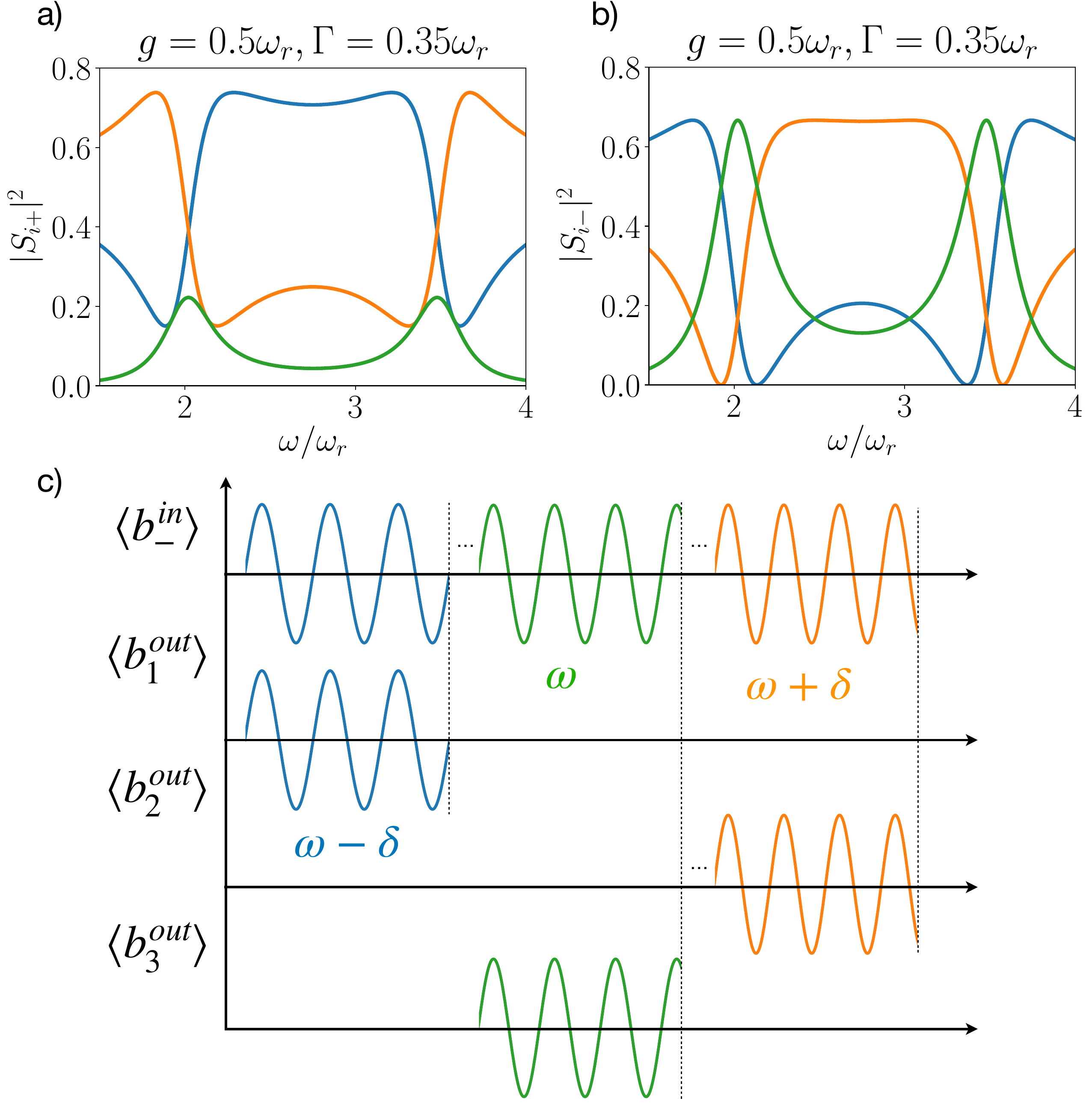}
\caption{(a,b) Outgoing power at each transmission line. The scattering matrix elements are  $S_{i\pm}=b_i^{out}/b_{\pm}^{in}$ when the input modes are given by $b_{\pm}^{in}$, with $i=1,2,3$ modes represented by the blue, orange and green curves respectively. The coupling strength to the resonators is set to $g=0.5 \omega _{r}$ and the effective photon decay rate is set to $\Gamma = 0.35 \omega _{r}$, see Appendix \ref{sec: scattering matrix} for the dependence of the scattering matrix elements on the circuit parameters. (c) Input-output scheme for three input signals with close frequencies. By choosing an effective photon decay rate and slightly tuning the input modes frequency, we can modulate the distribution of the output power through the transmission lines, the system behaving as a tunable directional coupler.}
\label{fig: outgoing powers}
\end{figure}

Finally, to check for the response of the JJ ring when we capacitively couple to three transmission lines, we study the scattering matrix of the system. This scattering matrix relates the input modes of the semi-infinite transmission lines with the output modes as sketched in Fig. \ref{fig: circuit}. When the plaquette hosts a chiral state, we expect a nonreciprocal behaviour between the input and the output modes. Taking the input-output relations for the transmission line modes \cite{Time reversal,Mueller_2018}
\begin{equation}
b_{j}^{out} \cor{\omega} =b_{j}^{in} \cor{\omega} + \frac{\Gamma}{3} \sum _{k=-1}^{1} \sum _{j'=1}^{3} \frac{e^{2 \pi i \parent{j-j'}k/3}}{i \parent{\omega - \Omega _{k}} - \frac{\Gamma}{2} } b_{j'}^{in} \cor{\omega}
\label{eq: input-output relations}
\end{equation}
with $\Gamma$ being the effective photon decay rate and $\Omega _{k}$ the frequencies of Eq. \eqref{eq: Omegas}. 

The complete S-matrix can be written as
\begin{equation}
\msf{S} = \begin{pmatrix}
\alpha & \beta e^{i2\pi/3} & \beta e^{-i2\pi/3} \\
\beta e^{-i2\pi/3} & \alpha & \beta e^{i2\pi/3} \\
\beta e^{i2\pi/3} &\beta e^{-i2\pi/3} & \alpha
\end{pmatrix}.
\end{equation}
with two complex parameters $\alpha$ and $\beta$ (see Appendix \ref{sec: scattering matrix} for more details).  The first thing to notice is that the S-matrix that relates the input and output modes, ${\bf{b^{out}}} = {\msf{S}} {\bf{b^{in}}}$, is not time-reversal symmetric, i.e., $\msf{S} \neq \msf{S}^{T}$. In fact, there is a nonreciprocal phase difference  $\arg{\left(S \right)}- \arg{\left( S^{T} \right)}=\frac{4 \pi}{3}$ for any value of the frequency $\omega/\omega_{r}$, coupling $g/\omega_{r}$, and decay $\Gamma/\omega_{r}$. Also $S S^{\dagger} = S^{\dagger} S = \mathbbm{1}$ as it should be by unitarity. Moreover, the output power in each transmission line is shown in Fig. \ref{fig: outgoing powers} for input modes $b_{+}^{in} = \frac{1}{\sqrt{2}}\parent{b_1^{in}+ b_2^{in}}$ and $b_{-}^{in} = \frac{1}{\sqrt{2}}\parent{b_1^{in}- b_2^{in}}$, i.e. $\tilde{S}=SU^{\dagger}$, where $U$ is the change of basis in the input ports. It is relevant to point out that by shifting the frequency of the input modes, most of the output power can be concentrated in any of the three ports at will with a maximal directionality of $2/3$, working as a tunable directional coupler. Due to the rotational symmetry of the setup, this behaviour is independent of the pair of continuous ports used for the input signal. Thus, the device can be seen as a circulator between differential input modes and local output modes.
 
\section{Conclusions}
In summary, we have shown the possibility of encoding chiral states in a superconducting Josephson junction plaquette that break time-reversal and parity symmetry. We have described a method to load these states in the proposed setup based on a spectral flow protocol, and we have discussed a possible way to access the non-trivial phases. Finally, we have analysed how such plaquettes can potentially become a fundamental unit of quantum nonreciprocal devices.

\begin{acknowledgments}
We thank the valuable and constructive comments by S. Girvin, C. M\"uller, and P. Zoller. Also, the authors acknowledge support from the projects QMiCS (820505) and OpenSuperQ (820363) of the EU Flagship on Quantum Technologies, Spanish MINECO/FEDER FIS2015- 69983-P, Basque Government IT986-16, EU FET Open Grant Quromorphic, and Shanghai STCSM (Grant No. 2019SHZDZX01-ZX04). This material is also based upon work supported by the U.S. Department of Energy.
\end{acknowledgments}

\appendix

\section{Circuit QED architecture: time dependant magnetic flux}

Following \cite{Time reversal,Mueller_2018}, we consider a minimal cQED set up consisting of a ring of three Josephson junctions threaded by an external magnetic flux with capacitive bias to ground, see Fig. \ref{fig1}. By making use of flux variable description \cite{Devoret_1995_QFluct,You_2019}, we define a set of branch variables $\bsb{\Psi}=\left(\Psi_1,\Psi_2,\Psi_3,\Psi_{J1},\Psi_{J2},\Psi_{J3}\right)$. Given that there are 3 main loops and one external flux, we will transform that set to $\bsb{\Phi}=\left(\Phi_1,\Phi_2,\Phi_3\right)$, through transformation 
\begin{eqnarray}
	\bsb{\Phi}&=&\msf{M}\bsb{\Psi},\\
	\msf{M}&=&\begin{pmatrix}
		\mathbbm{1}&0
	\end{pmatrix},
\end{eqnarray}
and $\mathbbm{1}$ is the 3-rank identity matrix. The external flux constraint is written as 
\begin{eqnarray}
	\bsb{\Phi}_e&=&(0,0,\phi_e)^T=\msf{R}\bsb{\Psi},\quad
	\msf{R}=\begin{pmatrix}
		\tilde{\msf{R}}&\mathbbm{1}\\
	\end{pmatrix},\\
	\tilde{\msf{R}}&=&\begin{pmatrix}
			1 & -1 & 0 \\
			0 & 1 & -1 \\
			-1 & 0 & 1
	\end{pmatrix}.
\end{eqnarray} 
We can remove the constraints of the original set through $\bsb{\Psi}=\msf{M}_+^{-1}\bsb{\Phi}_+$ where 
\begin{eqnarray}
	\bsb{\Phi}_+&=&\begin{pmatrix}
		\bsb{\Phi}\\
		\bsb{\Phi}_e
	\end{pmatrix}, \quad \msf{M}_+=\begin{pmatrix}
		\msf{M}\\
		\msf{R}
	\end{pmatrix}, \quad
	\msf{M}_+^{-1}=\begin{pmatrix}
		\mathbbm{1}&0\\
		-\tilde{\msf{R}}&\mathbbm{1}
	\end{pmatrix}.
\end{eqnarray}  
The Lagrangian of the system can be written and transformed as 
\begin{eqnarray}
	L&=&\frac{1}{2}\dot{\bsb{\Psi}}^T\mcl{C}\dot{\bsb{\Psi}}+\sum_i E_{Ji}\cos(2\pi\psi_i/\Phi_0)\nonumber\\
	&=& \frac{1}{2}\dot{\bsb{\Phi}}_+^T(\msf{M}_+^{-1})^T\mcl{C}\msf{M}_+^{-1}\dot{\bsb{\Phi}}_++\sum_i E_{Ji}\cos(\Delta\phi_i-\phi_{e,i})\nonumber\\
	&=&\frac{1}{2}\left(\dot{\bsb{\Phi}}^T\msf{C}\dot{\bsb{\Phi}}-\dot{\bsb{\Phi}}^T\tilde{\msf{R}}^T\msf{C}_J\dot{\bsb{\Phi}}_e-\dot{\bsb{\Phi}}_e^T\msf{C}_J\tilde{\msf{R}}\dot{\bsb{\Phi}}+\dot{\bsb{\Phi}}_e^T\msf{C}_J\dot{\bsb{\Phi}}_e\right)\nonumber\\
	&&+\sum_i E_{Ji}\cos(\Delta\phi_i-\phi_{e,i})
\end{eqnarray}
where the vector of phase variables are defined through the second Josephson relation $\phi_x=2\pi\Phi_x/\Phi_0$ with $\Phi_0$ the flux quantum constant. The complete capacitance matrix $\mcl{C}=\text{diag}(\msf{C}_G,\msf{C}_J)$, being $\msf{C}_G=C_G\mathbbm{1}$ and $\msf{C}_J=C_J\mathbbm{1}$. The differences of phases correspond to $\Delta\phi_i=\phi_{i+1}-\phi_i$, for $i=\{1,2,3\}$ and identifying $i=4$ with $i=1$. The capacitance matrix is defined as 
\begin{equation}
	\msf{C}=\msf{C}_G+\tilde{\msf{R}}^T\msf{C}_J\tilde{\msf{R}}=\begin{pmatrix}
		C_{\Sigma}&-C_J&-C_J\\
		-C_J&C_{\Sigma}&-C_J\\
		-C_J&-C_J&C_{\Sigma}\\	
\end{pmatrix},
\end{equation}
with $C_{\Sigma}=C_G+2C_J$. The (time-dependent) external flux appears naturally only in one cosine, i.e., for our specific choice of tree-branch variables $\varphi_{e,3}=\varphi_{e}(t)$ and zero otherwise. A linear transformation can be done to distribute this term in the potential at the expense of having a linear coupling of the time-dependent flux fluctuations with the fluxes in the kinetic energy \cite{You_2019},
\begin{equation}
	\bsb{\Phi}\rightarrow\bsb{\Phi}+\tilde{\msf{R}}^T\bsb{\Phi}_{e}/3\label{eq:ext_flux_transform}
\end{equation}
The Lagrangian is thus rewritten as, 
\begin{eqnarray}
	L&=& \frac{1}{2}\left(\dot{\bsb{\Phi}}+\tilde{\msf{R}}^T\dot{\bsb{\Phi}}_{e}/3\right)^T\msf{C}\left(\dot{\bsb{\Phi}}+\tilde{\msf{R}}^T\dot{\bsb{\Phi}}_{e}/3\right)\nonumber\\
	&&-\frac{1}{2}\left(\dot{\bsb{\Phi}}^T\tilde{\msf{R}}^T\msf{C}_J\dot{\bsb{\Phi}}_e+\dot{\bsb{\Phi}}_e^T\msf{C}_J\tilde{\msf{R}}\dot{\bsb{\Phi}}\right)\nonumber\\
	&&+\sum_i E_{Ji}\cos(\Delta\phi_i-\phi_{e}/3),
\end{eqnarray}
where we have removed the terms $\propto \dot{\Phi}_e^2$ as they will not give dynamics. The Legendre transformation involves the definition of conjugated charge variables $\bsb{Q}=\msf{C}(\dot{\bsb{\Phi}}+(\tilde{\msf{R}}^T/3-\msf{C}^{-1}\tilde{\msf{R}}^T\msf{C}_J)\dot{\bsb{\Phi}}_{e})$, to derive the Hamiltonian
\begin{eqnarray}
	H&=& \frac{1}{2}\bsb{Q}^T\left(\msf{C}^{-1}\bsb{Q}-2(\tilde{\msf{R}}^T/3-\msf{C}^{-1}\tilde{\msf{R}}^T\msf{C}_J)\dot{\bsb{\Phi}}_{e}\right)\nonumber\\
	&&-\sum_i E_{Ji}\cos(\Delta\phi_i-\phi_{e}/3),\label{eq:H_plaquette_ext_flux_app}
\end{eqnarray}
where again we have removed the terms $\propto\dot{\phi}_e^2$. It must be noted that a more general transformation can be performed in (\ref{eq:ext_flux_transform}) to annihilate the linear coupling for any time-dependent external flux, however, this will lead to the appearance of uneven external fluxes in the different cosines, see \cite{You_2019} for a full discussion on the choice of an {\it irrotational} constraint. For the sake of simplicity, let us work with the number of Cooper pair variables $\bsb{n}=\bsb{Q}/2e$, where $e$ is the electron charge. Conjugated classical variables are promoted to operators, with commutation relations $[n_i,  e^{\mp i\varphi_j}]=\mp\delta_{ij}e^{\mp i\varphi_j}$.

We recall that Hamiltonian (\ref{eq:H_plaquette_ext_flux_app}) has an important symmetry, readily, the total charge in the plaquette $N=\sum_i n_i$ is conserved quantity. We can perform a canonical transformation $\bsb{\varphi}\rightarrow\msf{T}\bsb{\varphi}$ and $\bsb{n}\rightarrow(\msf{T}^T)^{-1}\bsb{n}$, 
\begin{equation}
	\msf{T}=\begin{pmatrix}
		1 & 0 & 1 \\
		0 & -1 & 1 \\
		0 & 0 & 1 \\
	\end{pmatrix}
\end{equation}
to arrive to Hamiltonian
\begin{eqnarray}
	H&=&4E_\Sigma(\bsb{n}^T\msf{M}^{-1}\bsb{n})-\bsb{n}^T\msf{R}_e\dot{\bsb{\Phi}}_{e}-V(\bsb{\varphi},\phi_{e})\nonumber\\
	&=&4E_\Sigma\left(n_1^2+n_2^2+(n_2-n_1)n_3 -n_1n_2\right)\\
	&&+4E_3n_3^2-V(\bsb{\varphi},\phi_{e})\label{eq:H_plaquette_ext_flux_2},\\
	V&=&-E_{J2}\cos(\varphi_2-\phi_{e}/3)-E_{J3}\cos(\varphi_1-\phi_{e}/3)\nonumber\\
	&&	-E_{J1}\cos(\varphi_1+\varphi_2+\phi_{e}/3)
\end{eqnarray}
where  $E_\Sigma= e^2(\msf{C}^{-1}_{11}-\msf{C}^{-1}_{12})$ and $E_3=e^2\msf{C}^{-1}_{11}/2$, the inverse kinetic matrix is $\msf{M}^{-1}=(4E_\Sigma)^{-1}\msf{T}^{-1}\msf{C}^{-1}(\msf{T}^T)^{-1}$ and the flux coupling matrix is $\msf{R}_e=(4e)(\tilde{\msf{R}}^T/3-\msf{C}^{-1}\tilde{\msf{R}}^T\msf{C}_J)$.

\section{Continuum limit}
\label{sec: numeric estimation}
\begin{figure}[!]
\begin{minipage}[!]{\linewidth}
\includegraphics[width=\linewidth]{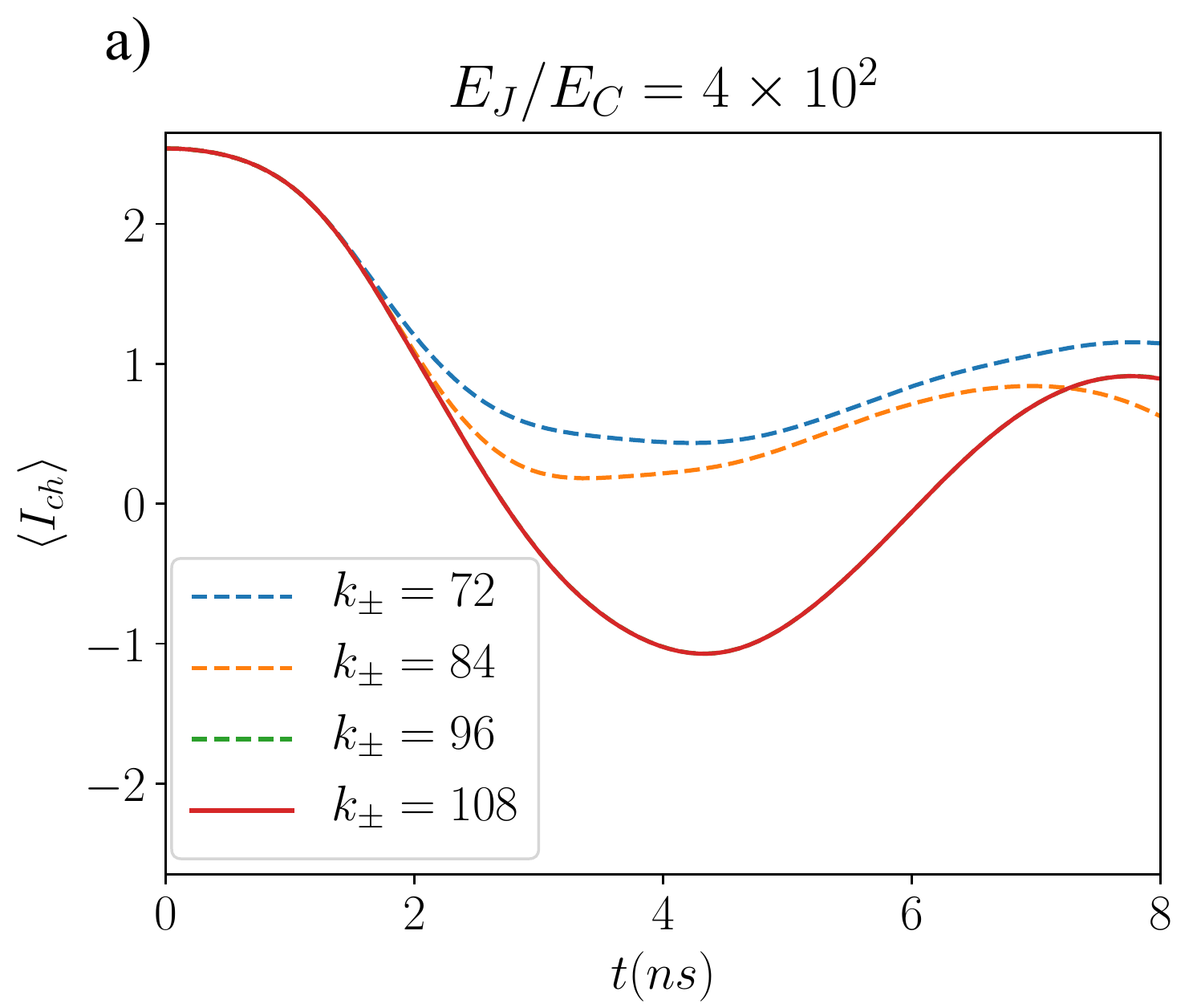}
\end{minipage}\vfill%
\begin{minipage}[!]{\linewidth}
\includegraphics[width=\linewidth]{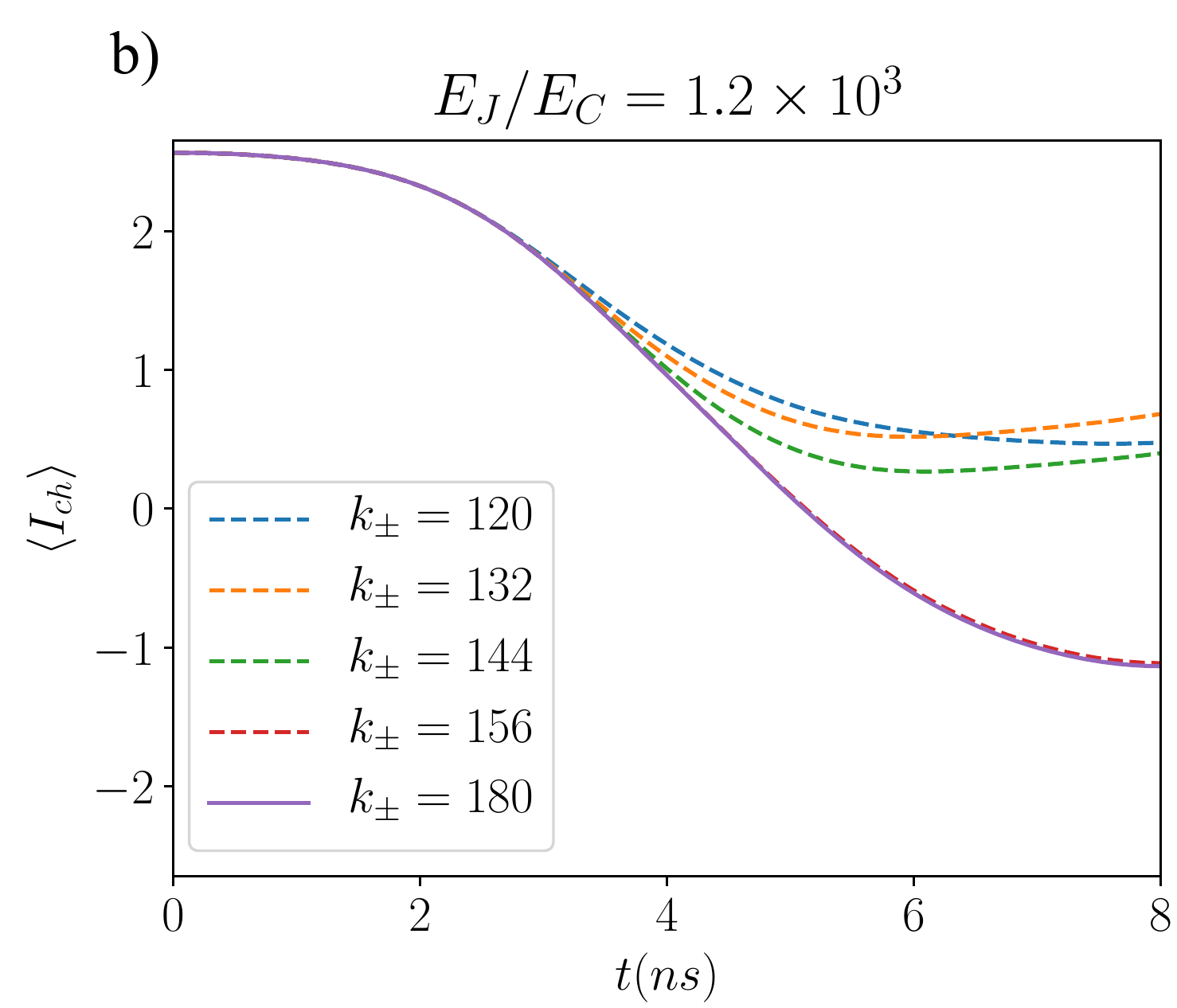}
\end{minipage}%
\caption{(a) Chiral current expected value time evolution and continuum limit with no approximations made. (b) The same plot for the biggest $E_{J}/E_{C}$ ratio we achieve with the machine.}
\label{fig: chiral current collapse}
\end{figure}

\begin{figure}[!]
\includegraphics[scale=0.55]{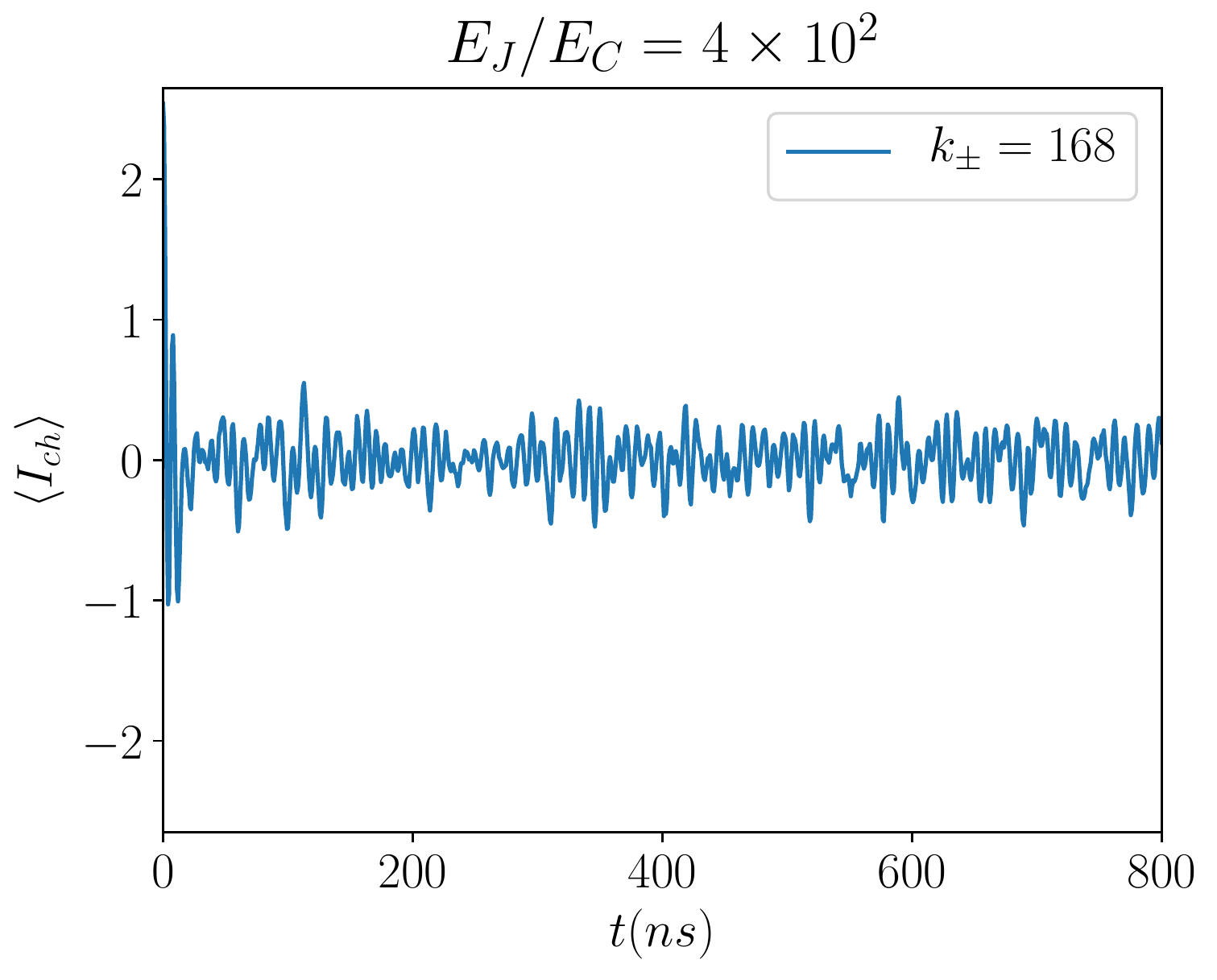}
\caption{Long time behaviour of the chiral current expected value for $E_{J}/E_{C} = 4 \times 10^{2}$ where no revivals are seen.}
\label{longtime}
\end{figure}

We have performed a numerical calculation for computing the chiral current. In doing so, we have discretised the Hilbert space and checked the correct behaviour in the continuum limit. Phase and charge basis are related by a Fourier transform $\bra{k_{\pm}} n_{\pm} \rangle = \frac{1}{\sqrt{L}} e^{i \frac{2 \pi}{L}k_{\pm} n_{\pm}}$, where $\ket{k_{\pm}}$ are eigenstates of the phase operators and $\ket{n_{\pm}}$ are eigenstates of the charge operators. In the phase basis the kinetic energy of the system \eqref{eq: harmonic Hamiltonian}
\begin{equation*}
\begin{split}
T &= \frac{E_{C}}{2L} \sum _{k_{\pm},\tilde{k}_{\pm},n_{\pm}} \left[3 \parent{n_+ - \frac{2}{3}N}^2 e^{i \frac{2 \pi}{L} \parent{k_+-\tilde{k}_+}n_+} \ket{k_+} \bra{\tilde{k}_+} \right. \\
& + \left. n_-^2 e^{i \frac{2 \pi}{L} \parent{k_- -\tilde{k}_-}n_-} \ket{k_-} \bra{\tilde{k} _-} \right],
\end{split}
\end{equation*}
with $\{k_{\pm}, \tilde{k} _{\pm},n_{\pm} \} \in \left[ - \frac{L}{2}+1, \frac{L}{2} \right]$ and the potential energy
\begin{equation*}
\begin{split}
V &= - E_J \sum _{k_{\pm}} \left[2 \cos \parent{\frac{2 \pi k_+}{L}}\cos \parent{\frac{2 \pi k_-}{L}   - \frac{\phi_{e}}{3}} \right.\\
& \left.+ \cos \parent{\frac{4 \pi k_-}{L}  + \frac{\phi_{e}}{3}} \right]  \ket{k_+, k_-} \bra{k_+,k _-} .
\end{split}
\end{equation*}
The curves of Fig. \ref{fig: chiral current collapse} correctly collapse for different EJ/EC ratios. The main conclusion of this analysis is that the charge energy $E_{C}$ is a relevant perturbation, giving a finite life-time to the chiral states. Furthermore, we have checked long time limit in Fig. \ref{longtime} where there is no signature of revivals of these states.

Next, we consider the harmonic Hamiltonian of the plaquette because it allows to estimate the number of free charges required to faithfully simulate the ground state of this Hamiltonian. Let's take the oscillator displaced by the external flux in the Hamiltonian \eqref{eq: harmonic Hamiltonian}
\begin{equation}
\begin{split}
H_{-} & = E_{C}n_{-}^{2} + 3 E_{J} \parent{\phi _{-} + \frac{\phi_{e}}{3}}^{2} \\
& = \frac{1}{2} \omega \parent{b^{\dagger}b + 1},
\end{split}
\label{eq: flux hamiltonian}
\end{equation}
with frequency $\omega = \sqrt{12 E_{J} E_{C}}$. The ground state of this Hamiltonian is a coherent state of the zero flux Hamiltonian, such that $b^{\dagger} = a^{\dagger} - \alpha, b = a - \alpha$ where $a \ket{\alpha} = \alpha \ket{\alpha}$ is the new coherent state of the zero-flux Hamiltonian and  
\begin{equation}
\phi _{-} = \parent{\frac{E_{C}}{3E_{J}}}^{\frac{1}{4}} \parent{a + a^{\dagger}},
\end{equation}
\begin{equation}
n _{-} = i \parent{\frac{3E_{J}}{E_{C}}}^{\frac{1}{4}} \parent{a - a^{\dagger}}.
\end{equation}
Introducing these definitions in Hamiltonian \eqref{eq: flux hamiltonian}
\begin{equation}
H = \omega \left[ \parent{a^{\dagger} + \alpha} \parent{a+ \alpha} + \frac{1}{2} \right]  + \frac{\phi_{e}^{2} E_{J}}{3} ,
\end{equation}
with $\alpha =\frac{\phi_{e}}{\sqrt{3}} \parent{\frac{E_{C}}{3E_{J}}}^{\frac{1}{4}}$. The expected value of the number operator in this coherent state $\langle n \rangle _{\alpha}$
\begin{equation}
 \bra{\alpha} a^{\dagger} a \ket{\alpha} = \vert \alpha \vert ^{2} = \frac{\phi_{e}^{2}}{3} \parent{\frac{E_{C}}{3E_{J}}}^{\frac{1}{2}}
\end{equation}
Therefore, we need at least $\langle n \rangle _{\alpha}$ excitations in this basis in order to properly represent the chiral state that the plaquette needs to host. Within this limit, in the harmonic approximation we expect the ground state wave-function of the $2 \pi$ flux Hamiltonian to oscillate with a frequency of the order of $\sqrt{12 E_{J}E_{C}}$. Indeed, this is the behaviour we observed, as it is shown in Fig. \ref{fig: chiral currents}. By increasing the $E_{J}/E_{C}$ ratio by a factor of four, the period of the oscillation doubles.

\section{Hamiltonian of resonators}
\label{appendix: transition probabilities}

To derive the resonators' effective Hamiltonian we use a second order Schrieffer-Wolff transformation. In the first place we take the ring to be in the state $\ket{N, \frac{2 \pi}{3}, -\frac{2 \pi}{3}}$ and the resonators being in a state with an arbitrary number of excitations $\ket{n_{a},n_{b},n_{c}}$. We label the state of the whole system by $\ket{\Psi _{i}}$, and impose that the plaquette remains with the same number of excitations $N$ after each second order transition. These transitions take the circuit from $\ket{\Psi _{i}}$ to $\ket{\Psi _{f}} = \ket{N, \frac{2 \pi}{3}, -\frac{2 \pi}{3}, n'_{a},n'_{b},n'_{c}}$. The matrix elements of the effective Hamiltonian are given by 

\begin{equation}
\frac{1}{2} \sum _{k,\alpha} C_{k, \alpha} \bra{\Psi _{i}} H_{\text{int}} \ket{\chi _{k,\alpha}} \bra{\chi _{k,\alpha}} H_{\text{int}} \ket{\Psi _{f}},
\end{equation}
where $C_{k, \alpha} = \frac{1}{E_{i}-E_{k,\alpha}}+\frac{1}{E_{f}-E_{k,\alpha}}$ being $E_{i},E_{f},E_{k,\alpha}$ the initial, final and intermediate energies of the circuit and $\ket{\chi _{k,\alpha}} = \ket{N',\varphi _{2}, \varphi _{3},n''_{a},n''_{b},n''_{c}}$ are the intermediate states with $k$ labelling the state of the ring and $\alpha$ the state of the resonators. The interaction Hamiltonian is the one appearing in Eq. \eqref{eq: Interaction Hamiltonian}. Taking into account that the exponentials of the phases act on the charge basis states as creation and annihilation operators 

\begin{equation}
\begin{split}
& \frac{1}{4} E_{J}^{r} \sum _{k,\alpha} C_{k, \alpha} \bra{\Psi _{i}} H_{\text{int}} \ket{\chi _{k,\alpha}} \bra{\chi _{k,\alpha}} \left(  \right. \\
& \ket{N+1, \varphi, n'_{a}-1,n'_{b},n'_{c}} + e^{-i \frac{2 \pi}{3}} \ket{N+1, \varphi, n'_{a},n'_{b}-1,n'_{c}} \\
& + e^{i \frac{2 \pi}{3}} \ket{N+1, \varphi, n'_{a},n'_{b},n'_{c}-1} + \ket{N-1, \varphi, n'_{a}+1,n'_{b},n'_{c}}  \\
& + e^{i \frac{2 \pi}{3}} \ket{N-1, \varphi, n'_{a},n'_{b}+1,n'_{c}}   \\
& + e^{-i \frac{2 \pi}{3}} \ket{N-1, \varphi, n'_{a},n'_{b},n'_{c}+1} \left. \right] = \\
& g \sum _{n'_{a},n'_{b},n'_{c}} \bra{\Psi _{i}} N,\frac{2 \pi}{3}, -\frac{2 \pi}{3} \rangle \left( 3 \ket{n'_{a},n'_{b},n'_{c}}  \right. \\
& + e^{i \frac{2 \pi}{3}} \ket{n'_{a}-1,n'_{b}+1,n'_{c}} + e^{-i \frac{2 \pi}{3}} \ket{n'_{a}-1,n'_{b},n'_{c}+1} \\
& + e^{-i \frac{2 \pi}{3}} \ket{n'_{a}+1,n'_{b}-1,n'_{c}} + e^{i \frac{2 \pi}{3}} \ket{n'_{a}+1,n'_{b},n'_{c}-1} \\
& + e^{-i \frac{2 \pi}{3}} \ket{n'_{a},n'_{b}+1,n'_{c}-1} + e^{i \frac{2 \pi}{3}} \ket{n'_{a}-1,n'_{b}-1,n'_{c}+1} \left. \right) \\
& = \sum _{i=a,b,c} 3g a_{i}^{\dagger}a_{i} \\
&+ \frac{g}{2} \left( a_{b}^{\dagger} a_{a} e^{-i \frac{2\pi}{3}} + a_{c}^{\dagger} a_{b} e^{-i \frac{2\pi}{3}}  + a_{a}^{\dagger} a_{c} e^{-i \frac{2\pi}{3}} + \text{h.c.} \right) .
\end{split}
\end{equation}
where we have restricted ourselves to the single excitation subspace in the resonators, and as $E_{i} = E_{f}$ then
\begin{equation}
g = \frac{1}{8} \parent{E_{J}^{r}}^{2} \left( \frac{2}{E_{i}-E_{N+1}}+\frac{2}{E_{i}-E_{N-1}} \right).
\end{equation}

The energies are given by 
\begin{equation}
\begin{split}
E_{i} & = E_{f} = E_{N}N^{2} + \omega _{r} - 3E_{J} \cos \frac{2 \pi}{3}, \\
E_{N+1} & = E_{N} \parent{N+1}^{2} - 3E_{J} \cos \frac{2 \pi}{3}, \\
E_{N-1} & = E_{N} \parent{N-1}^{2} + 2 \omega _{r} - 3E_{J} \cos \frac{2 \pi}{3}.
\end{split}
\end{equation}
Therefore
\begin{equation}
g = \frac{1}{2} \parent{E_{J}^r}^{2}  \left[E_{N} \parent{1- \parent{\frac{\omega _{r}}{E_{N} }-2N}^{2}} \right]^{-1}.
\end{equation}

\begin{figure}[!]
\begin{minipage}[h]{\linewidth}
\includegraphics[width=.9\linewidth]{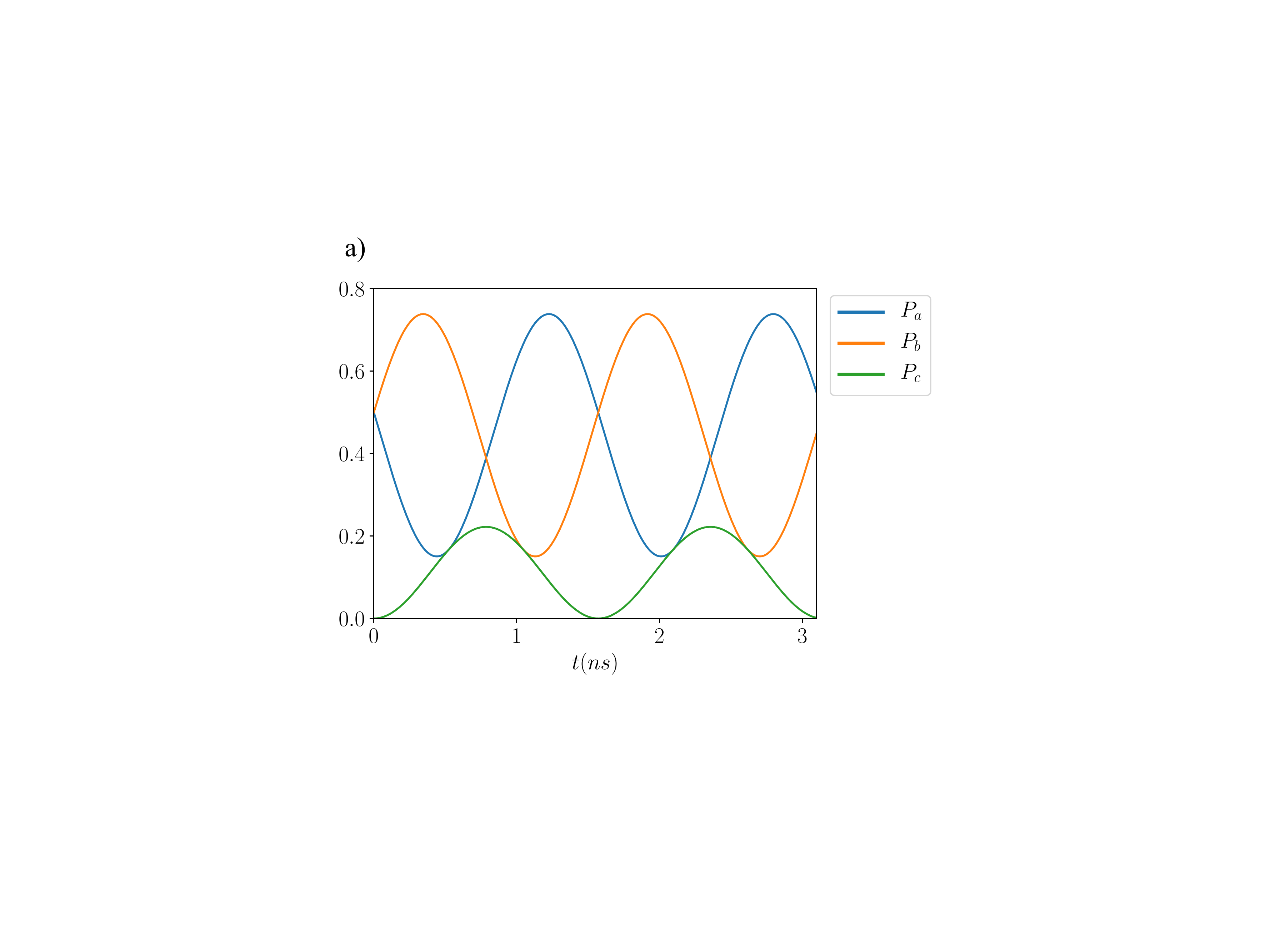}
\end{minipage}\vfill%
\begin{minipage}[h]{\linewidth}
\includegraphics[width=.9\linewidth]{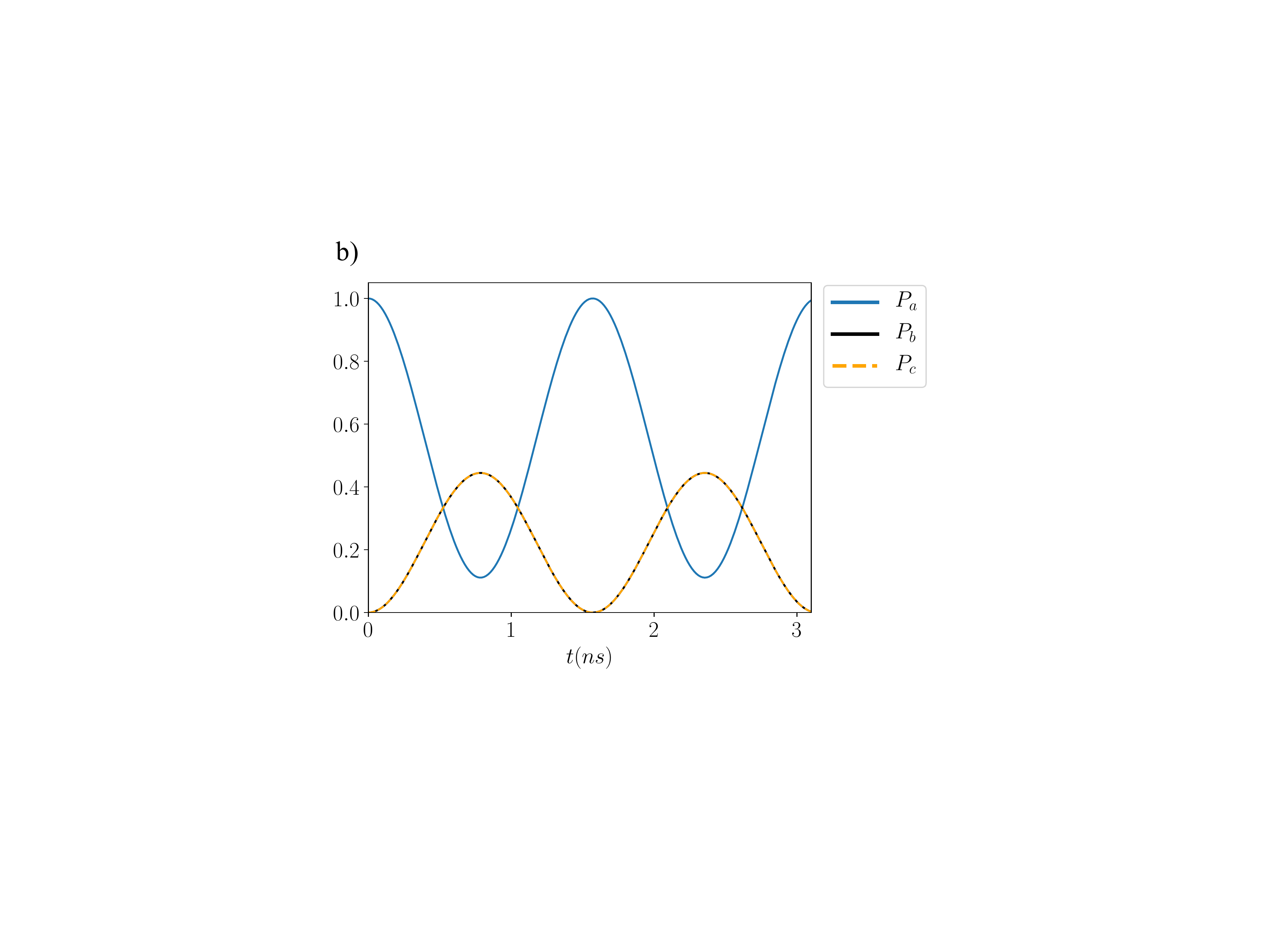}
\end{minipage}%
\caption{(a) Probability of finding the excitation at each resonator when the initial state of the resonators is  $\frac{1}{\sqrt{2}} \parent{\ket{100}+\ket{010}}$. (b) The same probabilities for the initial state being $\ket{100}$. Finding the excitation at the right hand side resonator or left hand side resonator is equally probable.}
\label{fig: resonators hopping 2}
\end{figure}

The chirality of the state in the plaquette has been made apparent by the circulation found in the resonators coupled through Josephson junctions to the plaquette with the initial state of the resonators being $\frac{1}{\sqrt{2}}\parent{\ket{100}- \ket{010}}$. However, in the basis we have chosen, it is also interesting to consider the hopping of the excitation through the resonators when the initial states are $\frac{1}{\sqrt{2}}\parent{\ket{100}+ \ket{010}},$ and $\ket{100}$. Fig. \ref{fig: resonators hopping 2} shows how the excitation finds no preferred direction to move when we place it in one resonator. An intermediate situation takes place in the remaining case.

\section{Scattering matrix}
\label{sec: scattering matrix}
The behaviour of the resonators under the introduction of different initial states through the transmission lines is given by the input-output relations for the modes of the lines of Eq. \eqref{eq: input-output relations}. The scattering matrix derived from the effective Hamiltonian \eqref{eq: effective Hamiltonian}
\begin{equation}
\msf{S} =\mathbbm{1} + \sum_{k} f_{k} \begin{pmatrix}
1 & e^{- i \frac{2 \pi}{3}k} & e^{i \frac{2 \pi}{3}k}\\
e^{i \frac{2 \pi}{3}k} & 1 & e^{- i \frac{2 \pi}{3}k}\\
e^{- i \frac{2 \pi}{3}k} & e^{ i \frac{2 \pi}{3}k} & 1
\end{pmatrix},
\end{equation}
where $f_k \parent{\omega}= \frac{\Gamma/3}{i \parent{\omega - \Omega _k} - \frac{\Gamma}{2}}$, $\Gamma = 2 \pi \abs{p}^2 \rho$ being the effective photon decay rate, $\rho$ the density of states of the transmission line, $p$ the interaction strength between the resonators and the transmission lines and $\Omega _{k}= 3g +  \omega _{r}  + 2g \cos \parent{\frac{2 \pi k}{3}+ \frac{2\pi}{3}}$.

There are three different entries in the complete S-matrix: 
\begin{equation*}
\begin{split}
&(\text{i})~  \alpha \equiv 1 +\sum_{k} f_{k} = 1+  \sum_{k} \frac{\Gamma/3}{i \parent{\omega - \Omega _k} - \frac{\Gamma}{2}} \\
&= 1+  \sum_{k} \frac{\Gamma/3}{i \left[\omega - 3g -  \omega _{r}  - 2g \cos \parent{\frac{2 \pi k}{3}+ \frac{2\pi}{3}} \right] - \frac{\Gamma}{2}} \\
&=1+  \left[ \frac{\Gamma/3}{i \left( \omega - \omega_{r} - 5g \right) - \frac{\Gamma}{2}} + \frac{2 \Gamma /3}{i \left( \omega - \omega_{r} - 2g \right) - \frac{\Gamma}{2}} \right],
\end{split}
\end{equation*}

\begin{equation*}
\begin{split}
&(\text{ii})~ \beta e^{-i2\pi/3} \equiv \sum_{k}   f_{k} e ^{i2\pi k/3} =  \sum_{k} \frac{  \Gamma e ^{i2\pi k/3}/3}{i \parent{\omega - \Omega _k} - \frac{\Gamma}{2}} \\
&=  \sum_{k} \frac{\Gamma e ^{i2\pi k/3}/3}{i \left[\omega - 3g -  \omega _{r}  - 2g \cos \parent{\frac{2 \pi k}{3}+ \frac{2\pi}{3}} \right] - \frac{\Gamma}{2}} \\
&=\frac{\Gamma e^{-i2\pi/3} /3}{i \left( \omega - \omega_{r} - 5g \right) - \frac{\Gamma}{2}} - \frac{ \Gamma e^{-i2\pi/3} /3}{i \left( \omega - \omega_{r} - 2g \right) - \frac{\Gamma}{2}},
\end{split}
\end{equation*}
\begin{equation*}
\begin{split}
&(\text{iii})~ \beta e^{i2\pi/3} \equiv \sum_{k}   f_{k} e ^{-i2\pi k/3} =  \sum_{k} \frac{  \Gamma e ^{-i2\pi k/3}/3}{i \parent{\omega - \Omega _k} - \frac{\Gamma}{2}} \\
&=  \sum_{k} \frac{\Gamma e ^{-i2\pi k/3}/3}{i \left[\omega - 3g -  \omega _{r}  - 2g \cos \parent{\frac{2 \pi k}{3}+ \frac{2\pi}{3}} \right] - \frac{\Gamma}{2}} \\
&=\frac{\Gamma e^{i2\pi/3} /3}{i \left( \omega - \omega_{r} - 5g \right) - \frac{\Gamma}{2}} - \frac{ \Gamma e^{i2\pi/3} /3}{i \left( \omega - \omega_{r} - 2g \right) - \frac{\Gamma}{2}},
\end{split}
\end{equation*}
with $\beta \equiv \frac{\Gamma  /3}{i \left( \omega - \omega_{r} - 5g \right) - \frac{\Gamma}{2}} - \frac{ \Gamma  /3}{i \left( \omega - \omega_{r} - 2g \right) - \frac{\Gamma}{2}}$. So, the complete S-matrix can be written as
\begin{equation}
\msf{S} = \begin{pmatrix}
\alpha & \beta e^{i2\pi/3} & \beta e^{-i2\pi/3} \\
\beta e^{-i2\pi/3} & \alpha & \beta e^{i2\pi/3} \\
\beta e^{i2\pi/3} &\beta e^{-i2\pi/3} & \alpha
\end{pmatrix}.
\end{equation}

\begin{figure}[!]
\includegraphics[scale=0.25]{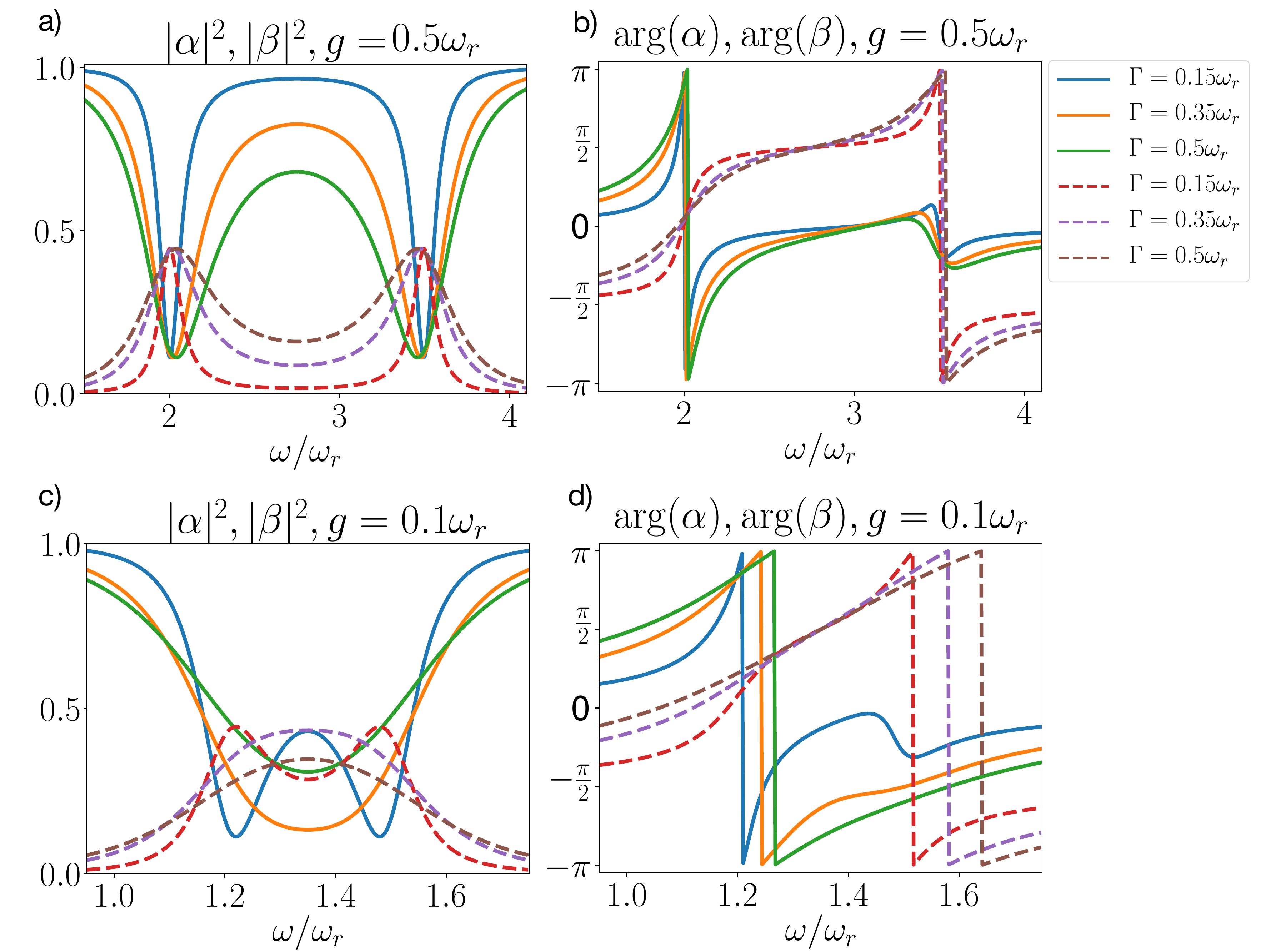}
\caption{Modulus (panel a) and c)) and argument (panel b) and d)) of the two parameters ($\alpha$ with continuous line, and $\beta$ with dashed line) that completely define the S-matrix of the studied system for different values of the frequency $\omega/\omega_{r}$, coupling $g/\omega_{r}$, and decay $\Gamma/\omega_{r}$. }
\end{figure}

The first thing to notice is that the S-matrix is not time-reversal symmetric which can be easily checked because $\msf{S} \neq \msf{S}^{T}$. In fact, there is a nonreciprocal phase difference  $\arg{\left(S \right)}- \arg{\left( S^{T} \right)}=\frac{4 \pi}{3}$ for any value of the frequency $\omega/\omega_{r}$, coupling $g/\omega_{r}$, and decay $\Gamma/\omega_{r}$. Also $S S^{\dagger} = S^{\dagger} S = \mathbbm{1}$ as it should be by unitarity.

Writing the scattering matrix in the basis generated by $\left\lbrace b_{+}^{\dagger},b_{-}^{\dagger},b_{3}^{\dagger}\right\rbrace$ for the input modes and going to the new input basis
\begin{equation}
\tilde{\msf{S}}=\msf{S}\msf{U}^\dag=\begin{pmatrix}
\frac{\alpha + \beta e^{i2\pi/3} }{\sqrt{2}} & \frac{\alpha - \beta e^{i2\pi/3} }{\sqrt{2}} & \beta e^{-i2 \pi /3} \\
\frac{\alpha + \beta e^{-i2\pi/3} }{\sqrt{2}} & \frac{ \beta e^{-i2\pi/3 } -\alpha }{\sqrt{2}} & \beta e^{i2 \pi /3} \\
-\frac{1}{\sqrt{2}} \beta& i \sqrt{\frac{3}{2}} \beta & \alpha
\end{pmatrix}.
\end{equation}

From the expressions of the S-matrix, it is straightforward to realise that $|\tilde{\msf{S}}_{3-}|^{2} = 3 |\tilde{\msf{S}}_{3+}|^{2} = \frac{3}{2} |\beta|^{2} = \frac{ 24 g^{2} \Gamma^{2}  }{ \left[  \Gamma^{2}  + 4  \left( \omega - \omega_{r} - 2 g \right)^{2} \right]  \left[  \Gamma^{2}  + 4  \left( \omega - \omega_{r} - 5 g \right)^{2} \right] }$ is the product of two Lorentzian distributions which is maximised at $\omega_{m} = \omega_{r} + 2 g$ and $\omega_{m} = \omega_{r} + 5 g$ with a value $|\tilde{\msf{S}}_{3-}|^{2}\Big|_{\omega \to \omega_{m}} \to \frac{24 g^{2}}{ \Gamma^{2} + \left( 6g \right)^{2}}$ and width proportional to $\Gamma$.

The new scattering matrix is non-symmetric and therefore, as long as the chiral state lives in the plaquette, the input-output scattering matrix holds and the effective model for the resonators is nonreciprocal. Indeed, in the limit of $E_{J} \gg E_{C}$ the state $\ket{N, \frac{2 \pi}{3}, - \frac{2 \pi}{3}}$ is an eigenstate of the ring Hamiltonian which implies the excitations living in the resonators can hop clockwise or counterclockwise indefinitely.
 
\begin{figure}[!]
\includegraphics[scale=0.25]{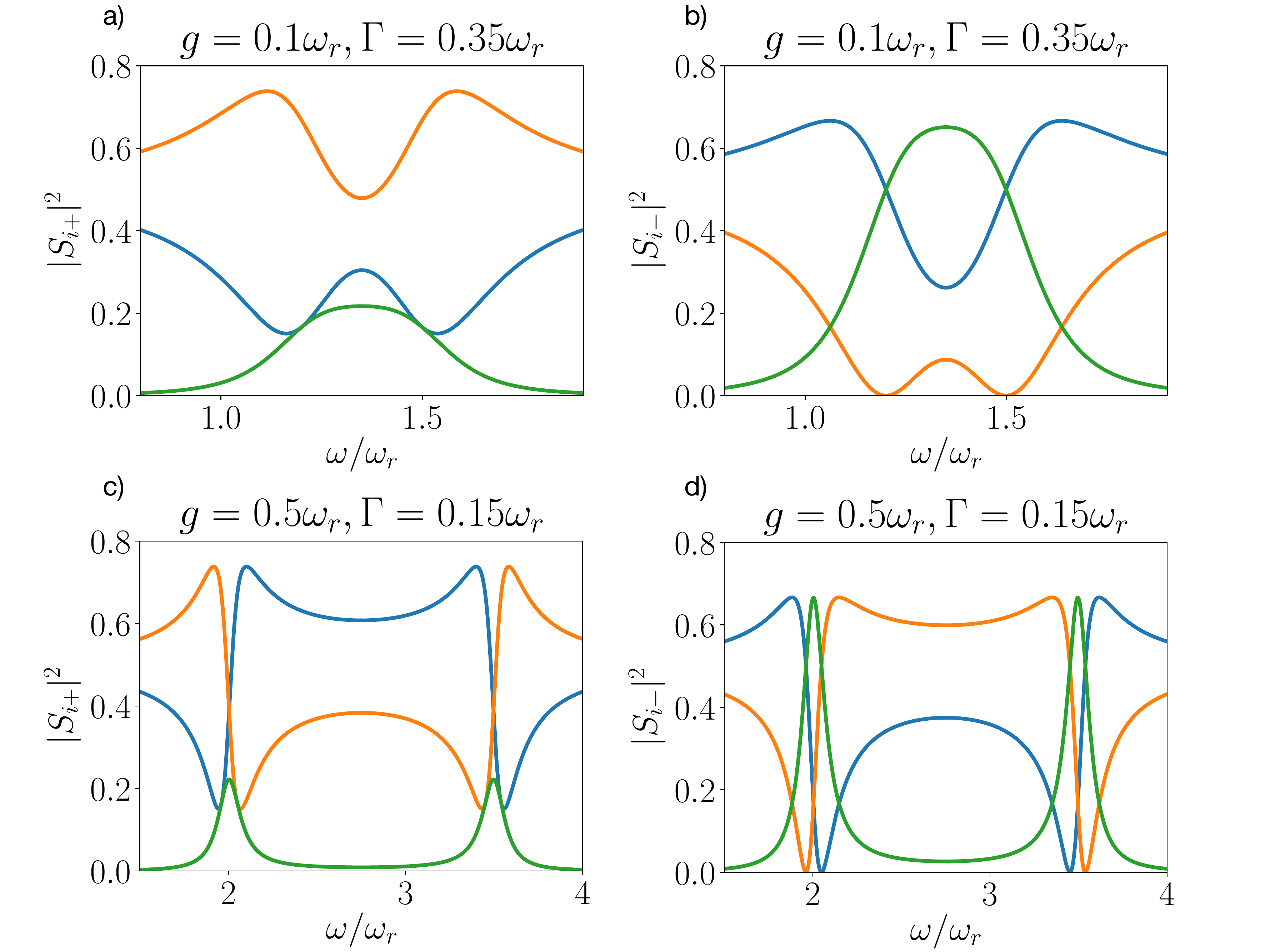}
\caption{The blue, orange and green lines represent scattering matrix elements corresponding to $i=1,2,3$ respectively. (a,b) Scattering matrix elements for $g > \Gamma$. (c,d) Same elements but with a smaller $\Gamma$ than shown in Section \ref{sec: preparation of the states}. Note that by choosing $g$ we can tune the interval in which the directional coupler behaviour takes place. The output in the third transmission line is described by a product of two Lorentzians centred in $\omega_{m} = \omega_{r} + 2 g$ and $\omega_{m} = \omega_{r} + 5 g$.}
\end{figure}

\section{Possible sources of disorder and decay}
\label{disorderanddecay}

The results shown in the main text require a long life-time of the chiral states $|N,\frac{2 \pi}{3},-\frac{2 \pi}{3} \rangle$ and $|N,-\frac{2 \pi}{3},\frac{2 \pi}{3} \rangle$ and a robust dynamics given by eq. (\ref{eq: Plaquette Hamiltonian})
\begin{equation*}
H=E_{N} N^{2} -   V \left(\varphi_{2} , \varphi_{3} \right)
\end{equation*}
with $E_{N}= \frac{2e^{2}C_{J}}{C_{0} \left(C_{0} + 3 C_{J} \right) } $, $V \left(\varphi_{2} , \varphi_{3} \right)= E_{J} \cos{\left( \varphi_{2} - \frac{\phi_{e}}{3} \right)} + E_{J} \cos{\left( \varphi_{3} - \varphi_{2} - \frac{\phi_{e}}{3} \right)} +  E_{J} \cos{\left( \varphi_{3} + \frac{\phi_{e}}{3} \right)}$, and where 
\begin{equation*}
|N,\varphi_{2},\varphi_{3} \rangle =\sum_{\{n_{2},n_{3}\} \in \mathbb{Z}} e^{-i\varphi_{2} n_{2}} e^{-i\varphi_{3} n_{3}} |N - n_{2} - n_{3} , n_{2} , n_{3} \rangle
\end{equation*}
are the eigenvectors.

A possible list of sources of disorder or imperfections is:

\paragraph*{Fluctuations in the external magnetic flux.-} Tuning the external magnetic flux is needed for the loading of the initial chiral state $|N,\frac{2 \pi}{3},-\frac{2 \pi}{3} \rangle$ or $|N,-\frac{2 \pi}{3},\frac{2 \pi}{3} \rangle$. This happens at a $\phi_{e}=\pm 2 \pi$. Nonetheless, there is a whole region in the external flux where the quantum states $|N,\frac{2 \pi}{3},-\frac{2 \pi}{3} \rangle$ or $|N,-\frac{2 \pi}{3},\frac{2 \pi}{3} \rangle$ are the ground state of the Hamiltonian. $ |N,\frac{2 \pi}{3},-\frac{2 \pi}{3} \rangle$ when $\pi < \phi_{e} < 3 \pi$ and $ |N,-\frac{2 \pi}{3},\frac{2 \pi}{3} \rangle$ when $-3 \pi < \phi_{e} < - \pi$. With energies $H\big|_{|N,-\frac{2 \pi}{3},\frac{2 \pi}{3} \rangle}=E_{N} N^{2}- 3 E_{J}  \cos{\left( \frac{\phi_{e}+ 2\pi}{3} \right)}$, and $H\big|_{|N,\frac{2 \pi}{3},-\frac{2 \pi}{3} \rangle}=E_{N} N^{2}- 3 E_{J}  \cos{\left( \frac{\phi_{e}- 2\pi}{3} \right)}$. A possible fluctuation around these values $\phi_{e}\pm 2 \pi = \delta \phi_{e}$ introduces a second order correction to the eigenvalues of the energies $\delta E = \frac{1}{6} E_{J} \left( \delta \phi_{e} \right)^{2}$.

\paragraph*{Disorder of the Josephson energies in the ring.-} In this case, we assume that the energy of the Josephson junctions are different for every junction, then $V \left(\varphi_{2} , \varphi_{3} \right)= E_{J_{1}} \cos{\left( \varphi_{2} - \frac{\phi_{e}}{3} \right)} + E_{J_{2}} \cos{\left( \varphi_{3} - \varphi_{2} - \frac{\phi_{e}}{3} \right)} +  E_{J_{3}} \cos{\left( \varphi_{3} + \frac{\phi_{e}}{3} \right)} $, with $E_{J_{i}} = E_{J} + \delta E_{J_{i}}$. It happens that the basis $|N,\varphi_{2},\varphi_{3} \rangle$ that diagonalises the unperturbed Hamiltonian, is also eigenbasis of every cosines term individually. This perturbation introduces fluctuations in the eigenvalues of the eigenstates but do not couple them, keeping the Hamiltonian diagonal.

\paragraph*{Disorder of the local charge energies in the ring.-} The energy scales studied in the main text are given by $E_{N} \gg  E_{J} \gg E_{C}$, where $E_{C}$ is the local charge energy scale. In this cases, we have seen that this perturbation is already relevant giving some finite life-time to the chiral states. A possible disorder on this perturbation, i.e., $E_{C_{i}} = E_{C} + \delta E_{C_{i}} $ will introduce a perturbation on top of the studied in the main text.

\paragraph*{Local charge decay.-} In this case, we assume an effective model given by a master equation in Lindblad form,
\begin{equation}
\dot{\rho} = \frac{-i}{\hbar} \left[ H , \rho \right] + \gamma \sum_{k=1,2,3} \left(  L_{k} \rho L_{k}^{\dagger} - \frac{1}{2} \left\{ L_{k}^{\dagger} L_{k} , \rho \right\} \right)
\end{equation}
where $H$ is the initial Hamiltonian, with damping rate $\gamma \ge 0$ and quantum jump operators $L_{k} = e^{-i\phi_{k}}$. Several properties are important to notice about the jump operators. They describe the local charge decay at every node $k$, i.e., $e^{-i \phi_{k}} | n_{k} \rangle = |n_{k} -1 \rangle$. Their action on the eigenstates $|N,\varphi_{2},\varphi_{3} \rangle$ of the Hamiltonian $H$ are: $e^{-i \phi_{1}} |N,\varphi_{2},\varphi_{3} \rangle = |N-1,\varphi_{2},\varphi_{3} \rangle $, $e^{-i \phi_{2}} |N,\varphi_{2},\varphi_{3} \rangle =e^{-i \varphi_{2}}   |N-1,\varphi_{2},\varphi_{3} \rangle $, and $e^{-i \phi_{3}} |N,\varphi_{2},\varphi_{3} \rangle =e^{-i \varphi_{3}}   |N-1,\varphi_{2},\varphi_{3} \rangle $. They are unitary operators $e^{-i \phi_{k}} e^{i \phi_{k}} = 1$. With these properties, it is easy to check that the set of steady states are given by a classical equal mixture of states with different $N$, i.e. $\rho_{t \to \infty} \left( \varphi_{2} , \varphi_{3} \right) = \sum_{N} |N,\varphi_{2},\varphi_{3} \rangle \langle N,\varphi_{2},\varphi_{3} |$. It is interesting to realise that the phase properties remains unchanged with these type of decay channel.


\begin{thebibliography}{}

\end{thebibliography}


\begin{thebibliography}{40}
\bibitem{Introduction to QEM circuits} U. Vool, M. H. Devoret, Int. J. Circ. Theor. Appl. {\bf45}, 897-934 (2017)
\bibitem{Microwave Photonics 2017} X. Gu, A. F. Kockum, A. Miranowicz, Y. X. Liu and F. Nori, Phys. Rep. {\bf718-719}, 1-102 (2017)
\bibitem{Frontiers of the Josephson effect} F. Tafuri, \textit{Fundamentals and Frontiers of the Josephson Effect} (Springer, 2019)
\bibitem{Fazio 2012} R. Fazio and G. Sch\"on, Ann. Phys. {\bf3-4}, 113-117 (2012)
\bibitem{Fazio 2001} R. Fazio. H. v.d. Zant, Phys. Rep. {\bf355}, 235-334 (2001)
\bibitem{Synthetic gauge fields} A. Nunnenkamp, J. Koch, S. M. Girvin, New J. Phys. {\bf13}, 095008 (2011)
\bibitem{Chiral ground state currents} P. Roushan et al., Nat. Phys. {\bf13}, 146-151 (2017)
\bibitem{Gauge potentials} H. Alaeian et al., Phys. Rev. A {\bf 99}, 053834 (2019)
\bibitem{Electromagnetic nonreciprocity} C. Caloz, A. Al\`u, S. Tretyakov, D. Sounas, K. Achouri, Z.-L Deck-L\'eger, Phys. Rev. Appl. {\bf10}, 047001 (2018)
\bibitem{A. Kamal thesis} A. Kamal, \textit{Nonreciprocity in active Josephson junction circuits}, PhD dissertation, Yale University, 2013
\bibitem{Microwave gyrator} C. L. Hogan, Bell Sys. Tech. J. {\bf31}, 1-31
\bibitem{A. Kamal 2011} A. Kamal, J. Clarke and M. H. Devoret, Nat. Phys. {\bf7}, 311-315 (2011)
\bibitem{Graph based analysis} L. Ranzani, J. Aumentado, New J. Phys. {\bf17}, 023024 (2015)
\bibitem{Quantum Hall circulator} A. C. Mahoney, J. I. Colless, S. J. Pauka, J. M. Hornibrook, J. D. Watson, G. C. Gardner, M. J. Manfra, A. C. Doherty, D. J. Reilly, Phys. Rev. X {\bf7}, 011007 (2017)
\bibitem{Self impedance circulator} S. Bosco, F. Haupt, D. P. DiVincenzo, Phys. Rev. Appl. {\bf7}, 024030 (2017)
\bibitem{Hall effect circulator} G. Viola, D. P. DiVincenzo, Phys. Rev. X {\bf4}, 021019 (2014)
\bibitem{Byeong 2012} Byeong Ho Eom, P. K. Day, H. G. LeDuc, J. Zmuidzinas, Nat. Phys. {\bf8}, 623-627 (2012)
\bibitem{near_quantum_limited_TWPA} C. Macklin, K. O Brien, D. Hover, M. E. Schwartz, V. Bolkhovsky, X. Zhang, W. D. Oliver, I. Siddiqi, Science {\bf350}, 307-310 (2015)
\bibitem{Flux driven JPA} T. Yamamoto, K. Inomata, M. Watanabe, K. Matsuba, T. Miyazaki, W. D. Oliver, Y. Nakamura, J. S. Tsai, Appl. Phys. Lett. {\bf 93}, 042510 (2008)
\bibitem{Low noise kinetic inductance} M. R. Vissers, R. P. Erickson, H.-S. Ku, L. Vale, X. Wu, G. C. Hilton and D. P. Pappas, Appl. Phys. Lett. {\bf 108}, 012601 (2016)
\bibitem{Optimizing Josephson ring modulator} C. Liu, T.-C. Chien, M. Hatridge, D. Pekker, Phys. Rev. A {\bf101}, 042323 (2020)
\bibitem{Widely tunable parametric amplifier} M. A. Castellanos-Beltra and K. W. Lehnert, Appl. Phys. Lett. {\bf91}, 083509 (2007)
\bibitem{M. H. Devoret 2010} N. Bergeal, R. Vijay, V. E. Manucharyan, I. Siddiqi, R. J. Schoelkopf, S. M. Girvin and M. H. Devoret, Nat. Phys. {\bf6}, 296-302 (2010)
\bibitem{Nonlinearities and parametric amp} E. A. Thol\'en, A. Erg\"ul, E. M. Doherty, F. M. Weber, F. Gr\`egis et al, Appl. Phys. Lett. {\bf90}, 253509 (2007)
\bibitem{A. Kamal 2013} B. Abdo, A. Kamal, M. H. Devoret, Phys. Rev. B {\bf87}, 014508 (2013)
\bibitem{Baleegh Abdo 2019} B. Abdo, N. T. Bronn, O. Jinka, S. Olivadese, A. D. C\'orcoles, V. P. Adiga, M. Brink, R. E. Lake, X. Wu, D. P. Pappas, J. M. Chow, Nat. Comm. {\bf10}, 3154 (2019)
\bibitem{Lecoc 2017} F. Lecocq, L. Ranzani, G. A. Peterson, K. Cicak, R. W. Simmonds, J. D. Teufel, J. Aumentado, Phys. Rev. Appl. {\bf7}, 024028 (2017)
\bibitem{Mechanical on-chip microwave} S. Barzanjeh, M. Wulf, M. Peruzzo, M. Kalaee, P. B. Dieterle, O. Painter and J. M. Fink, Nat. Comm. {\bf8}, 953 (2017)
\bibitem{Reconfigurable Josephson circulator} K. M. Sliwa, M. Hatridge, A. Narla, S. Shankar, L. Frunzio, R. J. Schoelkopf and M. H. Devoret, Phys. Rev. X {\bf5} 041020 (2015)
\bibitem{Nonreciprocal reconfigurable circuit} N.R. Bernier, L. D. T\'oth, A. Koottandavida, M. A. Ioannou, D. Malz, A. Nunnenkamp, A. K. Feofanov and T. J. Kippenberg, Nat. Comm. {\bf8}, 604 (2017)
\bibitem{Passive on-chip}  C. M\"uller, S. Guan, N. Vogt, J. H. Cole and T. M. Stace, Phys. Rev. Lett. {\bf120} 213602 (2018)
\bibitem{Kerckhoff 2015} J. Kerckhoff, K. Lalumi\`ere, B. J. Chapman, A. Blais and K. W. Lehnert, Phys. Rev. Appl. {\bf4} 034002 (2015)
\bibitem{ChiralSpinStates} X. G. Wen, F. Wilczek and A. Zee, Phys. Rev. B {\bf39}, 11413 (1989)
\bibitem{Quantum phase slips} G. Rastelli, I. M. Pop and F. W. J. Hekking, Phys. Rev. B {\bf87}, 174513 (2013)
\bibitem{Time reversal} J. Koch, A. Houck, K. Le Hur, S. M. Girvin, Phys. Rev. A {\bf82}, 043811 (2010).
\bibitem{Mueller_2018}C. Mueller, S. Guan, N. Vogt, J. H. Cole, and T. M. Stace, Phys. Rev. Lett. {\bf120}, 213602 (2018).
\bibitem{Devoret_1995_QFluct} M. H. Devoret, in \textit{Quantum Fluctuations in Electrical Circuits}, Proceedings of the Les Houches Summer School, Session LXIII (Elsevier Science B. V., New York, 1995).
\bibitem{Atom-Photon Interactions} C. Cohen-Tannoudji, J. Dupont-Roc and G. Grynberg, \textit{Atom-Photon Interactions: Basic Processes and Applications} 
\bibitem{You_2019}X. You, J. A. Sauls, and J. Koch Phys. Rev. B {\bf99}, 174512 (2019).
(Wiley, New York, 1998), Chap. B.
\bibitem{Introduction to quantum noise} A. A. Clerk, M. H. Devoret, S. M. Girvin, F. Marquardt and R. J. Schoelkopf, Revs. Mod. Phys. {\bf82}, 1155 (2010).
\end{thebibliography}
\end{document}